\documentclass[a4paper,11pt]{article}
\pdfoutput=1 

\usepackage{graphicx,color}
\usepackage{amsmath}
\usepackage{amssymb}
\usepackage{booktabs}
\usepackage[normalem]{ulem}
\usepackage[T1]{fontenc} 
\usepackage{subfig}
\usepackage{subfiles}
\usepackage[export]{adjustbox}
\usepackage{xcolor}
\usepackage{xspace}
\usepackage{euclid}
\usepackage[utf8]{inputenc}
\usepackage[switch, modulo]{lineno}
\usepackage{jcappub} 

\newcommand{\Om}{\Omega_\mathrm{m}\xspace}
\newcommand{\om}{\omega_\mathrm{m}\xspace}
\newcommand{\ob}{\omega_\mathrm{b}\xspace}
\newcommand{\rd}{\mathrm{d}}

\title{Probing the Distance Duality Relation with Machine Learning and Recent Data}

\author[a]{Felicitas Keil,}
\author[b]{Savvas Nesseris,}
\author[a, c, d]{Isaac Tutusaus,}
\author[a]{Alain Blanchard}
\affiliation[a]{Institut de Recherche en Astrophysique et Plan\'{e}tologie (IRAP), Universit\'{e} de Toulouse, CNRS, UPS, CNES, 14 Av.~Edouard Belin, 31400 Toulouse, France}
\affiliation[b]{Instituto de Fisica Teorica (IFT) UAM-CSIC, C/ Nicolas Cabrera 13-15, Campus de
Cantoblanco UAM, 28049 Madrid, Spain}
\affiliation[c]{Institute of Space Sciences (ICE, CSIC), Campus UAB, Carrer de Can Magrans, s/n, 08193 Barcelona, Spain}
\affiliation[d]{Institut d'Estudis Espacials de Catalunya (IEEC), Edifici RDIT, Campus UPC, 08860 Castelldefels (Barcelona), Spain}
\emailAdd{felicitas.keil@irap.omp.eu}
\emailAdd{savvas.nesseris@csic.es}
\abstract{
The distance duality relation (DDR) relates two independent ways of measuring cosmological distances, namely the angular diameter distance and the luminosity distance.
These can be measured with baryon acoustic oscillations (BAO) and Type Ia supernovae (SNe Ia), respectively. Here, we use recent DESI DR1, Pantheon+, SH0ES and DES-SN5YR data to test this fundamental relation. We employ a parametrised approach and also use model-independent Generic Algorithms (GA), which are a machine learning method where functions evolve loosely based on biological evolution. When we use DESI and Pantheon+ data without Cepheid calibration or big bang nucleosynthesis (BBN), there is a $2\sigma$ discrepancy with the DDR in the parametrised approach. Then, we add high-redshift BBN data and the low-redshift SH0ES Cepheid calibration. This reflects the Hubble tension since both data sets are in tension in the standard cosmological model $\Lambda$CDM. In this case, we find a significant violation of the DDR in the parametrised case at $6\sigma$. Replacing the Pantheon+ SNe Ia data by DES-SN5YR, we find similar results. For the model-independent approach, we find no deviation in the uncalibrated case and a small deviation with BBN and Cepheids which remains at 1$\sigma$. This shows the importance of considering model-independent approaches for the DDR.
}
\arxivnumber{2504.01750}

\begin{document}
\maketitle
\flushbottom

\section{Introduction}\label{sec:intro}

The Etherington \cite{etherington_lx_1933} or distance duality relation (DDR) relates the angular diameter distance with the luminosity distance. It holds in metric theories of gravity that have conservation of the photon number and where photons travel along unique null geodesics. Thus, testing this relation can rule out extensions to the standard cosmological model $\Lambda$CDM or it can be a strong indicator for new physics. Deviations from the DDR can be expressed with a function that quantifies this discrepancy. The DDR has been extensively tested and has gained renewed interest recently \cite{Alfano:2025gie, Yang:2025qdg, Gahlaut:2025lhv, Teixeira:2025czm, Wang:2025gus}. 

The DDR could be violated if there are deviations from a metric theory of gravity, if photons do not travel along null geodesics or if fundamental constants vary in time. The violation photon number conservation would break the DDR and could be caused, i.e. by attenuation through interstellar gas, dust or plasma \cite{Menard:2009yb}. Measuring the DDR can also provide limits on axion-like particles which change photon number as discussed in \cite{Avgoustidis:2010ju, Bassett:2003zw}. These are scalars beyond the standard model of particle physics that couple to photons. Thus, in presence of external magnetic fields, photons could convert into axion-like particles\cite{Raffelt:1987im}. The necessary conditions can possibly be provided by intergalactic magnetic fields.

Type Ia Supernovae (SNe Ia) allow for measurements of the luminosity distance when calibrated, e.g. with Cepheids. They can then be combined with other measurements related to the angular diameter distance. This has been done with measurements of the Hubble expansion rate from luminous red galaxies to constrain cosmic opacity and new physics \cite{Avgoustidis:2010ju}. Galaxy clusters can also be used for measurements on the angular diameter distance to compare them with SNe Ia data \cite{Holanda:2010vb, Cardone:2012vd, Holanda:2011hh, Goncalves:2011ha}. A common measurement of the angular diameter distance are baryon acoustic oscillations (BAO).  In \cite{Jesus:2024nrl}, Pantheon+ SNe Ia \cite{Scolnic:2021amr} and a compilation of BAO data (SDSS \cite{10.1093/mnras/stu523}, WiggleZ \cite{Blake:2012pj}, DES \cite{DES:2017rfo}) were used to test different parametrisations of DDR violations. 

Other than BAO, the DDR can also be tested using SNe Ia combined with strong gravitational lensing. No statistically significant violation has been found with the Pantheon+ data set \cite{Tang:2024zkc, Qi:2024acx}. This has been confirmed using compact radio quasars for the angular diameter distance, also in combination with Pantheon+ \cite{Yang:2024icv}. 

It is important to test the DDR in a model-independent way that does not assume a certain parametrisation that is not necessarily physically motivated. The authors of \cite{Xu:2022zlm} have conducted a model-independent test of the DDR utilising Pantheon SNe Ia \cite{Pan-STARRS1:2017jku} and eBOSS DR16 quasar data \cite{2021PhRvD.103h3533A}, similar to \cite{Wang:2024rxm} who use SDSS data \cite{SDSS:2000hjo}. In \cite{Wang:2025gus}, the authors use eBOSS and DESI in conjunction with DES-SN5YR and Pantheon+ data employing neural networks and finding a $2\sigma$ discrepancy at the higher redshifts in this sample.
BAO have also been used in \cite{Ma:2016bjt} with Monte Carlo methods to constrain the DDR violation agnostically at different redshifts. Strong gravitational lensing with SNe Ia has been tested model-independently as well, yielding no deviation \cite{Gahlaut:2025lhv}. Recent SNe Ia data was also used to reconstruct DDR deviations in a model-independent way and using only transversal BAO distances in \cite{Favale:2024sdq}. The authors do not use Cepheid calibration and find that the DDR holds in this case. However, in \cite{Alfano:2025gie}, data from the Sunyaev-Zeldovich effect for galaxy clusters are used, finding a $2$--$3$ $\sigma$ discrepancy depending on the cluster data. 

Here we use genetic algorithms (GA), which are machine learning methods. GA have already been used in other cosmological contexts, e.g. as a null test for the cosmological constant \cite{Nesseris:2010ep}, for the dark energy equation of state \cite{Bogdanos:2009ib, Nesseris:2012tt, Nesseris:2013bia} or other consistency tests including the Om statistic and the $r_0$ test for dark sector interactions \cite{Euclid:2021frk}. They also have been employed to reconstruct the Hubble expansion history \cite{Arjona:2019fwb} and to test curvature \cite{Sapone:2014nna} or the coupling of dark energy to the electromagnetic sector \cite{Euclid:2021cfn}.

The DDR has already been tested with GA using older data sets, e.g. in \cite{Nesseris:2012tt}. The authors in \cite{Euclid:2020ojp} used GA with Pantheon SNe Ia and various BAO surveys (6dFGS \cite{Beutler:2011hx}, SDSS \cite{10.1093/mnras/stu523}, BOSS CMASS \cite{Xu:2012hg}, WiggleZ \cite{Blake:2012pj}, MGS \cite{Ross:2014qpa}, BOSS DR12 \cite{BOSS:2015npt}, DES \cite{DES:2017rfo}, Ly-$\alpha$ \cite{eBOSS:2019qwo}, SDSS DR14 LRG \cite{eBOSS:2017tey} and quasar observations \cite{eBOSS:2017cqx}). So far, no studies have found significant violation.

Other models outside the DDR have been tested with DESI DR1 \cite{DESI:2024mwx} and Pantheon+ \cite{Scolnic:2021amr} and alternatively DES-SN5YR \cite{DES:2024jxu}. In \cite{Mukherjee:2024ryz}, the authors reconstruct the expansion history with the Om statistic and the evolution of the total equation of state, finding a $2 \sigma$ discrepancy for Pantheon+ and an over $3 \sigma$ discrepancy with DES-SN5YR data. An extension of the Friedman--Lemaître--Robertson--Walker (FLRW) metric has been tested on this data \cite{Fernandez-Garcia:2025vnb}, finding a discrepancy to $\Lambda$CDM greater than $1 \sigma$.
In \cite{Tutusaus:2023cms} and references therein, it is shown that imposing the low- and high-redshift normalisations through the Cepheids and BBN calibrations leads to a violation of the DDR.

This work provides a model-independent approach for the DDR in addition to a para\-metrised approach using new data sets, namely Pantheon+, SH0ES \cite{Scolnic:2021amr}, DES-SN5YR \cite{DES:2024jxu} and DESI DR1 \cite{DESI:2024mwx}. For the parametrised method, we minimise the $\chi^2$ statistic numerically in every point over all other parameters whose contours are not considered. In a next step, we use GA as our model-independent approach. It is model-independent in the sense that there are no parameters that are being determined. Instead, the GA finds a functional form that fits the data as well as possible, based on the $\chi^2$ statistic as well. This is useful since we do not know a priori in which way the DDR would be violated, if at all.

In addition to the uncalibrated SNe Ia and BAO analysis, we calibrate our data sets in such a way that they reflect the Hubble tension which exists in $\Lambda$CDM. For a summary of the Hubble tension, see e.g. \cite{Schoneberg:2021qvd}. To put the DDR into the context of this tension, we calibrate the SNe Ia with SH0ES yielding $H_0=73.5\pm 1.1 \, \kmsMpc$ \cite{Brout:2022vxf} and the DESI BAO with the big bang nucleosynthesis (BBN) value for $\ob$, resulting in $H_0=68.53\pm 0.80 \, \kmsMpc$ \cite{DESI:2024mwx} for the parametrised approach. The connection between the DDR and the Hubble tension has also been explored by \cite{Teixeira:2025czm} with different parametrisations of the DDR, two of which solve the Hubble tension together with the wCDM model for dark energy. Here, we use a different parametrisation that modifies the BAO observables instead of the SNe Ia observables and also employ a model-independent approach, in addition to the parametrised one.

In Sect. \ref{sec:DDR}, we introduce and derive the DDR. Sect. \ref{sec:data} describes the SNe Ia and BAO data sets we employ and details the Cepheid calibration. The parametrisation with the $\epsilon$ parameter is presented in Sect. \ref{sec:param} and the model-independent GA approach in Sect. \ref{sec:GA}. We show our results for both approaches in Sect. \ref{sec:results} before we summarise and conclude in Sect. \ref{sec:conclusion}.

\section{Distance  Duality Relation}\label{sec:DDR}
The luminosity distance $d_{\rm{L}}$, which can be measured by SNe Ia, compares flux $F$ and absolute luminosity L of an object.
\begin{equation}
    d_{\rm{L}}=\sqrt{\frac{\mathrm{L}}{4\,\pi \,F}}\:.
\end{equation}
If we assume a flat universe, we can express the flux with the comoving distance $d_\mathrm{M}$
\begin{equation}
    F=\frac{L}{4\,\pi\, a_0^2\, d_\mathrm{M}^2\, (1+z)^2}\:.
\end{equation}
Here, $a_0$ is the scale factor of the universe today. The factor $(1+z)^2$ comes from the fact that the photon energy is proportional to $\frac{1}{a(t)}\propto (1+z)$ coming from the dilation of the wavelength. Additionally, the rate at which photons arrive, which is measured by the luminosity, is also diluted by $\frac{1}{a(t)}$.
Using this, we obtain the luminosity distance as a function of the comoving distance
\begin{equation}
    d_{\rm{L}}=a_0\, (1+z)\,d_\mathrm{M}\:.
\end{equation}
On the other hand, the angular diameter distance can be obtained using trigonometry by comparing the proper length $l$ and the angular size $\theta$ of an object. To do this, we have to know the proper length of the object which is the case for BAO, making it a standard ruler.
\begin{equation}
    d_{\rm{A}}=\frac{l}{\theta}\:.
\end{equation}
In the FLRW metric, a proper length is equal to the product of the scale factor, the comoving distance and the angular size, yielding $l=a(t)\,d_\mathrm{M}\,\theta$ \cite{Weinberg:2008zzc}. Due to the expansion of the universe, distances are redshifted by the factor $1+z=\lambda(t_0)/\lambda(t)=a_0/a(t)$ where $\lambda$ is the physical size. This leads us to:
\begin{equation}
    \frac{l}{\theta}=a(t)\,d_\mathrm{M}= \frac{a_0}{(1+z)}\,d_\mathrm{M}\:.
\end{equation}
The DDR now follows from comparing both distance measures
\begin{equation}
    d_{\rm{A}}=\frac{d_{\rm{L}}}{(1+z)^2}\:.
\end{equation}
To test this empirically, we parametrise deviations from the DDR by defining the following phenomenological function
\begin{equation}
    \eta (z)=\frac{d_{\rm{L}}(z)}{d_{\rm{A}}(z)\,(1+z)^2}\:.
    \label{eta-definition-distances}
\end{equation}
Thus, if $\eta(z)=1$, there is no discrepancy, which serves as the null hypothesis.

\section{Data} \label{sec:data}


\subsection{Supernovae Type Ia}\label{subsec:SN data}

The Pantheon+ SNe Ia data set \cite{Scolnic:2021amr} contains 1701 Type Ia supernovae. We follow the Pantheon+ likelihood \cite{Brout:2022vxf} and compute the luminosity distance with the Hubble Diagram Redshift. This is then rescaled using the heliocentric redshift for the likelihood since the apparent magnitude $m$ depends on the heliocentric redshift. For the magnitude, we use the Tripp 1998 \cite{Tripp:1997wt} corrected magnitude.
The measured apparent magnitude is then compared with the theoretical one calculated from the luminosity distance:
\begin{equation}
    m_{\rm{th}}(z,\Om,H_0)=M_0+5\,\log_{10}\left[\frac{d_{\rm{L}}(z,\,\Omega_{\rm m},\,H_0)}{\mathrm{Mpc}}\right]+25\;.
\end{equation}
Thus, our $\chi^2$ statistic depends on the parameters $\Omega_\mathrm{m}$, $H_0$ and $M_0$. We calculate it with the inverse of the covariance matrix $C_{ij}$ \cite{Tegmark:1996bz}:
\begin{equation}
    \chi^2_{\rm{Panth}}=(m_{\rm{th}}-m_{\rm{data}})_iC_{ij}^{-1}(m_{\rm{th}}-m_{\rm{data}})_j\, . 
\end{equation}
The Pantheon+ likelihood, released with the data set and the cosmological constraints \cite{Brout:2022vxf}, excludes some of the SNIa. When using the Pantheon+ data set without SH0ES, a redshift cut ($z>0.01$) has to be made due to systematic errors in the data \cite{Brout:2022vxf}. In the calibrated case, SNIa below this redshift that are not calibrators are removed. Furthermore, in the uncalibrated case, we marginalise analytically over the absolute magnitude $M_0$ with the expression taken from Appendix C of \cite{SNLS:2011lii}.

\subsubsection{Cepheid Calibration}\label{subsubsec:Cali}

When using the Pantheon+ SNe Ia catalogue in conjunction with SH0ES, it is possible to calibrate all SNe Ia with Cepheids. In the cases where a Cepheid has been found in the same galaxy as the SNe Ia, the apparent magnitude will be calculated with the absolute distance from the Cepheid:
\begin{equation}
    m(z)=M_0+d_{\mathrm{abs}, \mathrm{SH0ES}}.
\end{equation}
Every SN Ia with the calibration flag has this associated apparent magnitude instead of the one using the luminosity distance. This calibration effectively fixes the absolute magnitude $M_0$ and thus sets the Hubble parameter $H_0$ to the combined Pantheon+ and SH0ES value of $73.5\pm 1.1 \, \kmsMpc$ \cite{Brout:2022vxf}. 

\subsection{Baryon Acoustic Oscillations}\label{subsec:BAO}
In the early universe, before the decoupling of baryons from photons, acoustic waves were propagating in the tightly coupled fluid with the same phase. After decoupling, at the drag epoch, these waves were frozen-in and are the basis for matter fluctuations which grow with the expansion of the universe. This depends on the sound horizon $r_{\rm s}$ at the drag redshift $z_{\rm d}$ whose length is theoretically well-predicted using early-time physics.
It is obtained by integrating over the sound speed $c_\mathrm{s}$ from the drag redshift shortly after decoupling up to infinity:

\begin{equation}
    r_{\rm d}=r_{\rm s}(z_{\rm d})=\int_{z_{\rm d}}^\infty \frac{c_\mathrm{s}(z)}{H(z)}\mathrm{d}z\;.
    \label{rd-def}
\end{equation}
We use the DESI DR1 data release \cite{DESI:2024mwx} with seven redshift bins from $0.1<z<4.2$ using the bright galaxy survey, luminous red galaxies, emission line galaxies, quasars and Lyman-$\alpha$ data. Since angular separation $\Delta \theta$ is measured, this yields a ratio of a distance divided by the sound horizon at drag epoch $r_{\rm d}$. This is calculated with the approximation by \cite{Brieden:2022heh} which we adopt as well. It assumes $\Lambda$CDM and standard early-time physics to estimate the $r_{\rm d}$ given the physical matter and baryon densities $\om$ and $\ob$ and the effective number of neutrinos $N_{\rm{eff}}$:
\begin{equation}
    r_{\rm d}=147.05\,\left(\frac{\om}{0.1432}\right)^{-0.23}\,\left(\frac{N_{\rm{eff}}}{3.04}\right)^{-0.1}\,\left(\frac{\ob}{0.02236}\right)^{-0.13}\,\mathrm{Mpc}\;.
    \label{r_d_equation_desi}
\end{equation}
For the effective number of neutrinos, we use $N_{\rm{eff}}=3.044$ \cite{Froustey:2020mcq} throughout this whole analysis.
\newline
The distance in each BAO bin is either the comoving distance $d_{\rm M}(z)$ and the Hubble distance $d_{\rm H}(z)$ or the angle-averaged distance $d_{\rm V}(z)$ in bins with low signal-to-noise ratio:

\begin{equation}
    d_{\rm V}(z)=\left[z\,d_{\rm M}(z)^2\,d_{\rm H}(z)\right]^{1/3}\;.
\end{equation}
In the $\Lambda$CDM case, the comoving distance is the angular diameter distance multiplied by $1+z$:
\begin{equation}
    d_{\rm M}=d_{\rm A}\,(1+z)\;.
\end{equation}

In the flat $\Lambda$CDM model, BAO constrain two cosmological parameters: the matter density $\Omega_{\rm m}$ and the product $r_{\rm d} \cdot h$, where $h$ is the dimensionless Hubble parameter $h=H_0/(100\;\rm{km}\;\rm{s}^{-1}\;\rm{Mpc}^{-1})$.

In practice, we need to compute the function $d_{\rm A}(z)$ containing the additional factor $\eta(z)$ from which we derive $d_{\rm M}(z)$ and $d_{\rm H}(z)$ and $d_{\rm V}(z)$. In the next step, the ratio is computed with $r_{\rm d}$. The $\chi^2$  for the BAO is computed analogously to the SNe Ia case, with its respective covariance matrix $\Tilde{C_{ij}}$.
\begin{equation}
    \chi^2_{\rm{DESI}}=(v_{\rm{th}}-v_{\rm{data}})_i\,\Tilde{C}_{ij}^{-1}\,(v_{\rm{th}}-v_{\rm{data}})_j\;,
\end{equation}
where $v_i=d_{X,i}/r_{\rm d}$. We use all DESI DR1 data points provided in Table 1 of \cite{DESI:2024mwx}, i.e. $d_{\rm M}/r_{\rm d}$ and $d_{\rm H}/r_{\rm d}$ or only $d_{\rm V}/r_{\rm d}$, depending on the redshift bin. 

Since we use profile likelihoods, when we profile a given parameter, we minimise the $\chi^2$ statistic over $r_{\rm d}\cdot h$ at each point of the profile.  This is done in the standard case without BBN. In the cases for which we want to break the degeneracy between $r_{\rm d}$ and $h$, we use Eq. \ref{r_d_equation_desi}. 

\section{Parametrisation with the $\epsilon$-Parameter}\label{sec:param}
Since the deviations grow larger at higher redshifts, we parameterise $\eta(z)$ with a constant exponential parameter $\epsilon$ which is common in the literature \cite{Avgoustidis:2010ju, Holanda:2012ia, Jesus:2024nrl}:

\begin{equation}
    \eta(z)=(1+z)^\epsilon.
    \label{eta-z-epsilon}
\end{equation}
We test this parametrisation of $\eta(z)$ using a profile likelihood approach. In every calculated point, the $\chi^2$ statistic is calculated using the data sets, minimising over all other parameters.
For the BAO data, we need to know the evolution of $H(z)$. For the GA, this is computed from the angular diameter distance $d_{\rm A}=d_{\rm L}/[(1+z)^2\eta(z)]$ where $d_{\rm L}$ and $\eta(z)$ are the resulting GA functions. To calculate $H(z)$ from the GA results, we write the luminosity distance as a redshift integral and multiply by $\eta(z)$:

\begin{equation}
    d_{\rm L}(z, \Om)=(1+z)\,\eta(z)\,\int_0^z\:\frac{c}{H(z', \Om)}\,\rd z'\:.
\end{equation}
Now, we can derive both sides to find $H(z)/H_0$:
\begin{equation}
     \frac{H(z,\Om)}{H_0}=\left[\frac{\rd}{\rd z}\left(\frac{d_{\rm L}(z,\Om)}{c/H_0 (1+z)\eta(z)}\,\right)\right]^{-1}\;.
     \label{H_z_from_dL}
\end{equation}
From these, we can derive the other distances needed for the BAO. The $\chi^2$ is independent of $H_0$ as discussed in \ref{subsec:BAO}. 

Due to the $\epsilon$-parametrisation, only one data set is rescaled with $\epsilon$. We choose BAO in this case, so the SNe Ia likelihood does not depend on $\epsilon$ here. The usual approach modifies the SNIa observables and leaves the BAO observables unchanged, see e.g. \cite{Euclid:2020ojp}. However, our phenomenological approach modifies all quantities that describe the BAO data, i.e. $d_{\rm A}, d_{\rm H}$ and $d_{\rm V}$. The relation between the comoving distance and $H(z)$ stays intact as can be seen in Eq. \ref{H_z_from_dL} which also implies that $\eta(z)$ is necessary to calculate it. To calculate the $\chi^2$ for the BAO, we fit $d_{\rm A}(\Om, \epsilon)$ to the data using the modified DDR:

\begin{equation}
    d_{\rm A}(z,\Om,\epsilon)=d_{\rm L}(z,\Om)\cdot(1+z)^{-2-\epsilon}.
    \label{epsilon-def}
\end{equation}

We compute a profile likelihood for $\Om$ and $\epsilon$ using Eq. \ref{epsilon-def}. This statistical problem could be equally addressed in a Bayesian framework or with profile likelihoods (see e.g. chapter 4 of \cite{Wolschin:2010skf}). We choose to use profile likelihoods for our analysis since it makes it easier to directly compare the $\chi^2$ values between $\Lambda$CDM and the DDR violation model with $\epsilon$. The data and model are also simple enough that the computing resources are sufficiently small to conduct this analysis on a personal computer. Thus, the speed-up gained by using Bayesian analysis would be negligible.

\subsection{BBN Baryon Abundance for BAO}\label{subsec:BBN}

The DESI data by itself cannot constrain $H_0$, as detailed in Sect. \ref{subsec:BAO}. However, we can use Eq. \ref{r_d_equation_desi} to compute the sound horizon at the drag redshift. This depends on the value of the baryon density $\ob=\Omega_\mathrm{b}\,h^2$, which can be computed from the BBN. Using data about primordial Deuterium and Helium abundances yields $\ob=0.02218\pm 0.00055$ \cite{Schoneberg:2024ifp}. When this is added as a fixed parameter to the BAO data from DESI, it fixes $r_{\rm d}$, meaning we can compute $H_0$. The uncertainty on $\ob$ is very small, therefore we assume that there is no significant correlation with $\Om$ or $\epsilon$. As can be seen in the constraints from the DESI DR1 data release, the information from BBN in practice does not change the value or the error of $\Om$, as can be seen in \cite{DESI:2024mwx} in Table 3.

In $\Lambda$CDM, DESI+BBN is not compatible with Pantheon+SH0ES due to the Hubble tension. In our model for the DDR violation, we test whether the BBN value is compatible with the rest of the data set, i.e. DESI+Pantheon+SH0ES. In this case, we find a best-fit $\ob$ value that is 1.8 $\sigma$ from the BBN value. This means that the data sets are compatible with the BBN value in the $\epsilon$-parametrisation and we can therefore combine the different data sets.

\subsection{SH0ES Magnitude for DES-SN5YR}

The DES-SN5YR data are not directly coupled to a Cepheid data set. So, to incorporate the SH0ES value, we calculate the absolute magnitude $M_{0,\mathrm{SH0ES}}$. To do this, we average over all $M_0$ values in the calibrated Pantheon+SH0ES catalogue, fixing $\Om=\Omega_{\mathrm{m}, \mathrm{DES}}$ \cite{DES:2024jxu} and $H_0=H_{0,\mathrm{SH0ES}}$ \cite{Brout:2022vxf}. We repeat this for adding and subtracting the error from Pantheon+SH0ES, i.e. $H_0=H_{0,\mathrm{SH0ES}} \pm \sigma_{H_0,\mathrm{SH0ES}} $, to calculate the error $\sigma_{M_0, \mathrm{SH0ES}}$. Using these values, we construct an additional $\chi^2$ for the value of $M_0$:

\begin{equation}
    \chi^2_\mathrm{SH0ES}(M_0)=\left(\frac{M_0-M_{0,\mathrm{SH0ES}}}{\sigma_{M_0, \mathrm{SH0ES}}}\right)^2\;.
    \label{chi2-sh0es-h0}
\end{equation}
This essentially calibrates the SNe Ia with the SH0ES absolute magnitude which in turn determines $H_0$.

\section{Goodness of fit}\label{subsec:goodness-of-fit}
To estimate whether the model fits the data well, we use a statistic following \cite{Shafer:2015kda}. We calculate the probability $P(\nu, \chi^2)$ with which a higher value of $\chi^2$ would randomly occur. This is the cumulative distribution function of the $\chi^2$-distribution. Here $\nu$ is the number of degrees of freedom, i.e. the amount of data points $N$ subtracted by the number of free model parameters $k$: $\nu=N-k$. This probability is calculated using the $\chi^2$ of the considered model:
\begin{equation}
    P(\nu, \chi^2)=\frac{\Gamma(\frac{\nu}{2},\frac{\chi^2}{2})}{\Gamma(\frac{\nu}{2})}\, .
\end{equation}
Here, $\Gamma(z,x)$ is the upper incomplete gamma function and $\Gamma(z)=\Gamma(z,0)$ is the complete gamma function.
\begin{equation}
    \Gamma(z,x)=\int  t^{z-1} e^{-t} dt \, .
\end{equation}
The higher the value of $P(\nu, \chi^2)$ is, the higher is the probability that one would obtain a worse value of $\chi^2$. This means that a value close to 1 is an indicator for a high goodness-of-fit and the further $P(\nu, \chi^2)$ is from 1, the worse the fit is.

\section{Genetic Algorithms}\label{sec:GA}
GA are a Machine Learning method that finds functions to fit a data set without previous parametrisation. Here, we will briefly outline how they function. For a more detailed introduction, we refer the reader to Ref. \cite{Bogdanos:2009ib}. 

To find the best-fitting function, GA have a random initial population of functions according to a specified grammar. We use polynomials here because they are differentiable and sufficient to describe our data. From the initial population, in each generation the functions are evaluated for how well they match the data using a $\chi^2$-function which is the sum of the SNe Ia and BAO $\chi^2$ statistic:
\begin{equation}
    \chi^2_\mathrm{tot}=\chi^2_\mathrm{DESI}+\chi^2_\mathrm{Pantheon+SH0ES}.
\end{equation}
\begin{figure}[h]
	\centering
    \subfloat[{Crossover operation}]{
        \vspace{1cm}
    	\includegraphics[width=.5\textwidth]{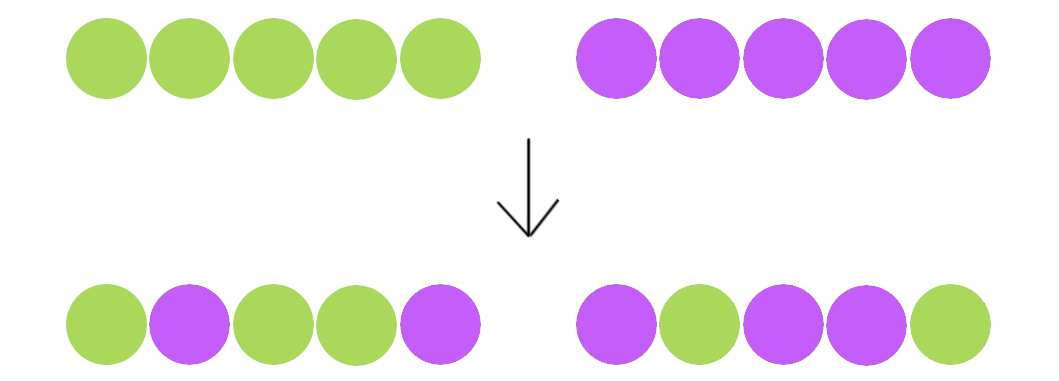}
    	\label{fig:crossover}}
    \hspace{1cm}
	\subfloat[{Mutation operation}]{
        \includegraphics[width=.21\textwidth]{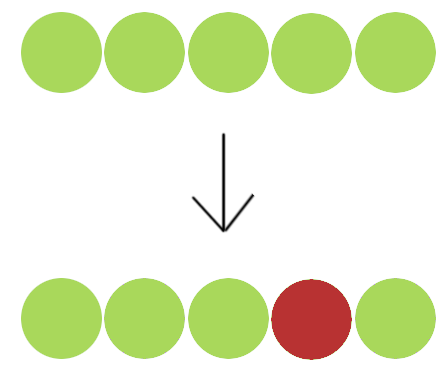}
    	\label{fig:mutation}
        \vspace{4cm}}
    \caption{Illustration of GA operations.}
\end{figure}
The functions with the lowest $\chi^2$ can generate offspring functions, and the rest do not propagate to the next generation. The remaining functions "reproduce", with a preselected probability called the crossover rate, see Fig. \ref{fig:crossover}. Here, both parent functions combine parts of them, which represents the DNA in the analogy. These parts are then summed. Analogously, the mutation operation (see Fig. \ref{fig:mutation}) is selected with a probability called the mutation rate. If that is the case, one parameter of the function will randomly change at the time of "reproduction" to the new generation. We repeat this for 10 samples. For each sample, the function with the lowest $\chi^2$ is the final candidate. From these 10, we select the one with the lowest $\chi^2$ as the resulting best-fit function.

We compare the resulting functions of the GA with $\Lambda$CDM for each case. When we calculate $H(z)$ in the $\Lambda$CDM case from the Friedmann equation, we also consider radiation and neutrinos.
\begin{equation}
    \frac{H^2(z)}{H_0^2}=a^{-3}\Om\left(1+a_{\rm{eq}}/a\right)+\left[1-\Om\left(1+a_{\rm{eq}}\right)\right]
\end{equation}

The scale factor at matter--radiation equality $a_{\rm{eq}}$ depends on the density of radiation $\Omega_\gamma$ and ultra-relativistic neutrinos $\Omega_\nu$:
\begin{equation}
    a_{\rm{eq}}=\frac{\Omega_\gamma+\Omega_\nu}{\Om}.
\end{equation}

The GA tries to fit two functions, namely the luminosity distance $d_{\rm L}$ and the DDR deviation $\eta(z)$, from which the angular distance is calculated. In the first three cases, we either marginalise analytically over the absolute SNe Ia magnitude $M_0$ or minimise the $\chi^2$ by varying $r_{\rm d}\cdot h$. This implies that the GA can rescale the distance functions by any value without changing the $\chi^2$. Any global factor is reabsorbed in the marginalisation or the minimisation, respectively. Thus, we choose a dimensionless reconstruction of the function in the first three cases.

We define a functional form that the GA follows which constrains the function space and ensures e.g. the boundary condition that $d_{\rm L}(0)=0$. This also leads to faster convergence. We have tested this without these constraints and obtained mostly unphysical results. The GA thus predicts the functions $f_\mathrm{GA}$ and $g_\mathrm{GA}$ in the following form:
\begin{equation}
    \eta(z)=(1+z)^{\epsilon(z)}=(1+z)^{1/10\,f_\mathrm{GA}(z)},
\end{equation}
\begin{equation}
    \frac{H_0}{c}\,d_{\rm L}(z)=z[1+z\, g^2_{\mathrm{GA}}(z)].
\end{equation}
Using these results, we can calculate $d_{\rm A}$ using Eq. \ref{eta-definition-distances} and $H(z)$ using Eq. \ref{H_z_from_dL}.
\begin{figure}
	\center
	\includegraphics[width=7cm]{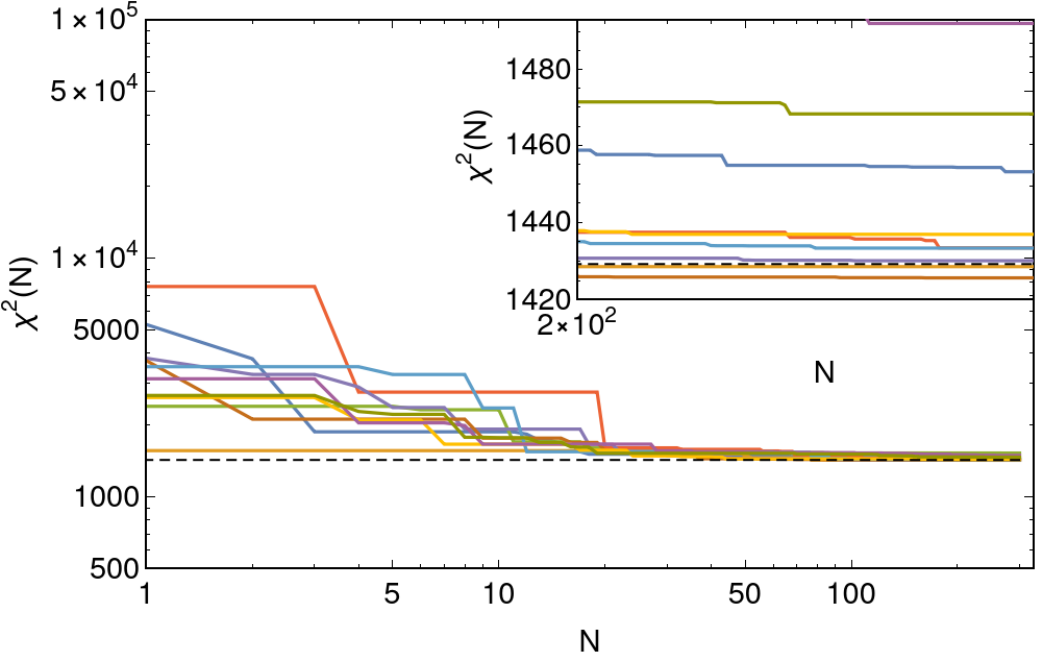}
	\caption{$\chi^2$ of the GA functions per generation for the standard DESI+Pantheon+ case.}
	\label{fig:chi2progression}
\end{figure}
In Fig. \ref{fig:chi2progression}, we show the evolution of the GA over 300 generations. All 10 populations approach the $\chi^2$ of $\Lambda$CDM for which we minimised over $H_0$ and $\Om$. This is represented by the dashed line. In this figure, we used the standard Pantheon+ case for which 5 of the 10 samples have a lower $\chi^2$ than $\Lambda$CDM at the end. 

For the model-independent GA, it is necessary to use the unparametrised sound horizon at the drag redshift $r_{\rm d}$. We cannot add only the BBN value into equation \ref{r_d_equation_desi} as we did in the parametrised case. This would imply feeding the GA a parametrised equation with cosmological parameters, even though the GA is unparametrised.  Thus, we calculate this quantity not only with BBN but with additional \textit{Planck} 2018 data. We then calculate a fixed value of the sound horizon using the \textit{Planck} value of $\omega_{\rm{m}}$ \cite{Planck:2018vyg} in addition to the BBN. Here, we could use both values for $\omega_{\rm{b}}$ and $\omega_{\rm{m}}$ from \textit{Planck}. However, since we only need $\omega_{\rm{m}}$ in the unparametrised case and prefer to use BBN to be CMB independent, we also use the BBN value for the GA for consistency reasons.

In the case using Cepheid calibration and the sound horizon with BBN and \textit{Planck} values, we do not predict the distances in a dimensionless way as before, since in this case it is possible for the GA to predict both the amplitude of $d_{\rm L}$ and $d_{\rm A}$. This emphasises the violation of the DDR due to the Hubble tension, since we predict $d_{\rm L}$ instead of $H_0\cdot d_{\rm L}$. However, it also increases the computing time for the GA since machine learning algorithms generally work better with values closer to unity \cite{Ioffe:2015ovl}. So, even after rescaling our functions, the GA needs 1000 generations in this case to converge compared to 300 in the standard case.

\subsection{Error Calculation}
To estimate the GA error, we generate multiple samples using different random seeds. This affects the initial population of the GA. From these samples of different resulting functions $\eta(z)$, we select the 68.27\% of functions closest to our best-fit solution. This would correspond to the 1$\sigma$ error in the Gaussian case, but we find that our results are not symmetric in every case. The uncertainty is then taken from 100 GA samples, calculated in redshift intervals of $0.1$ and then interpolated to smooth it out. We continually increased the number of samples and found that the error stabilises before or around 100 samples.
\section{Results}\label{sec:results}

\subsection{Parametrised Sampling}\label{subsec:parametrised-results}
\subsubsection{Pantheon+ with DESI DR1}\label{subsubsec:param-pantheon}

\begin{figure}[h]
    \centering
    \subfloat[{Contour plot for Pantheon+ (yellow), DESI (blue) and their combination (red).}
        ]{
        \includegraphics[width=7cm, clip]{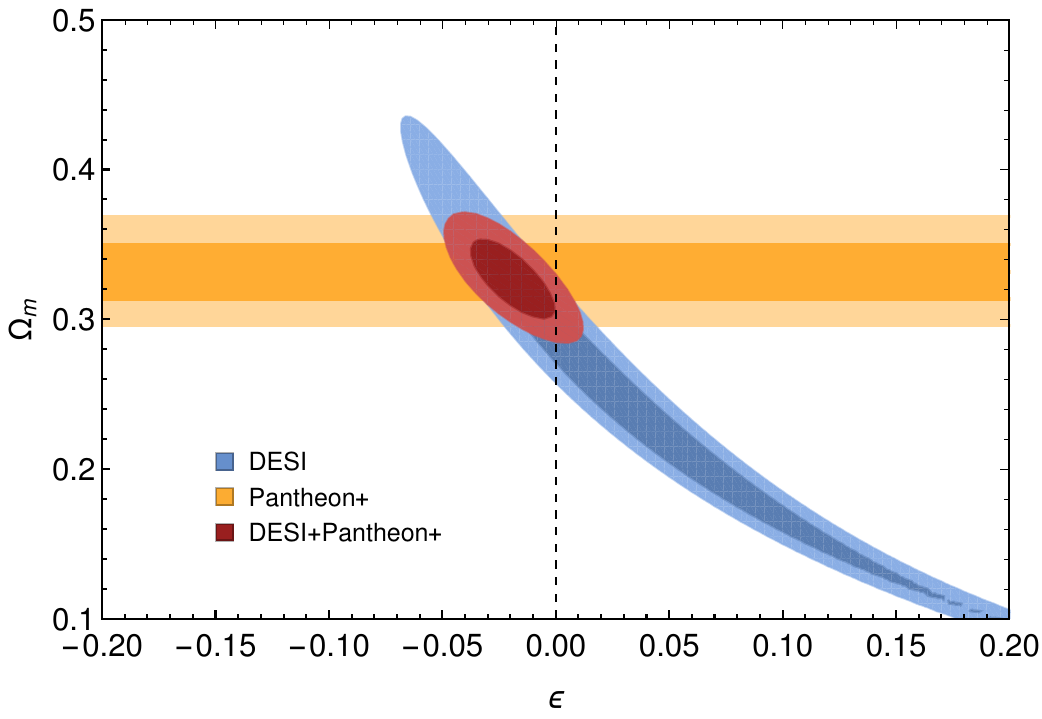}
        \label{fig:Om-eps-dL-dim}}
    \hspace{0.5cm}
    \subfloat[{Contour plot for Pantheon+ (yellow), DESI+BBN  (blue) and their combination (red).}
        ]{
        \includegraphics[width=7cm, clip]{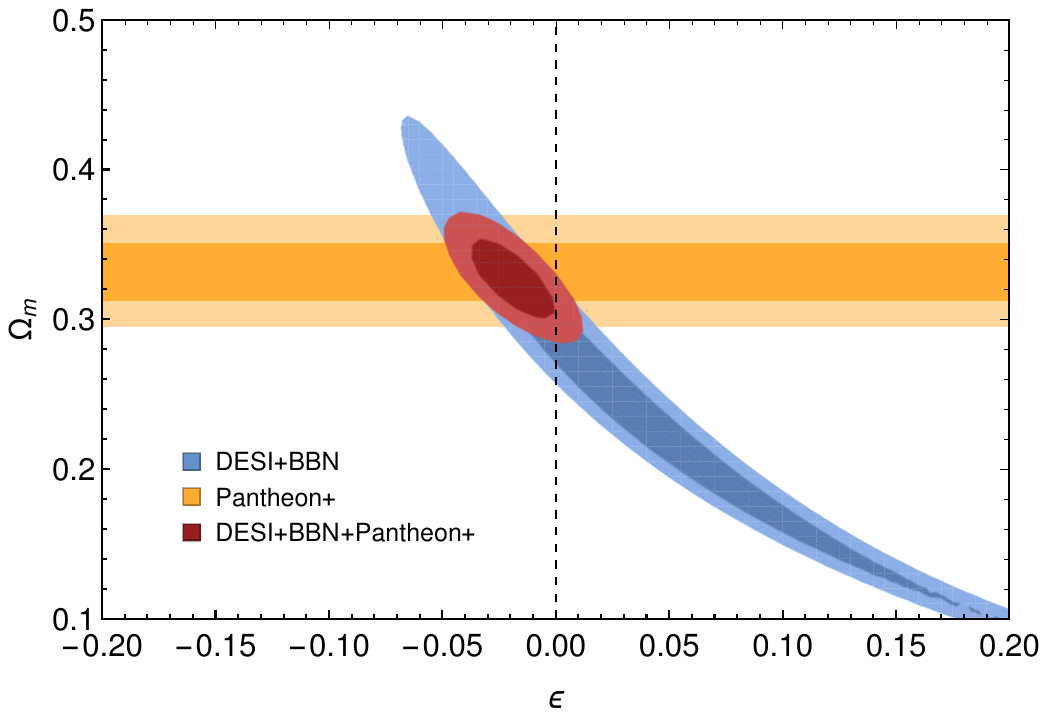}
        \label{fig:Om-eps-dL-dim-planck}}
    \\
    \subfloat[{Contour plot for Pantheon+SH0ES (yellow), DESI (blue) and their combination (red).}
    ]{
	   \includegraphics[width=7cm, clip]{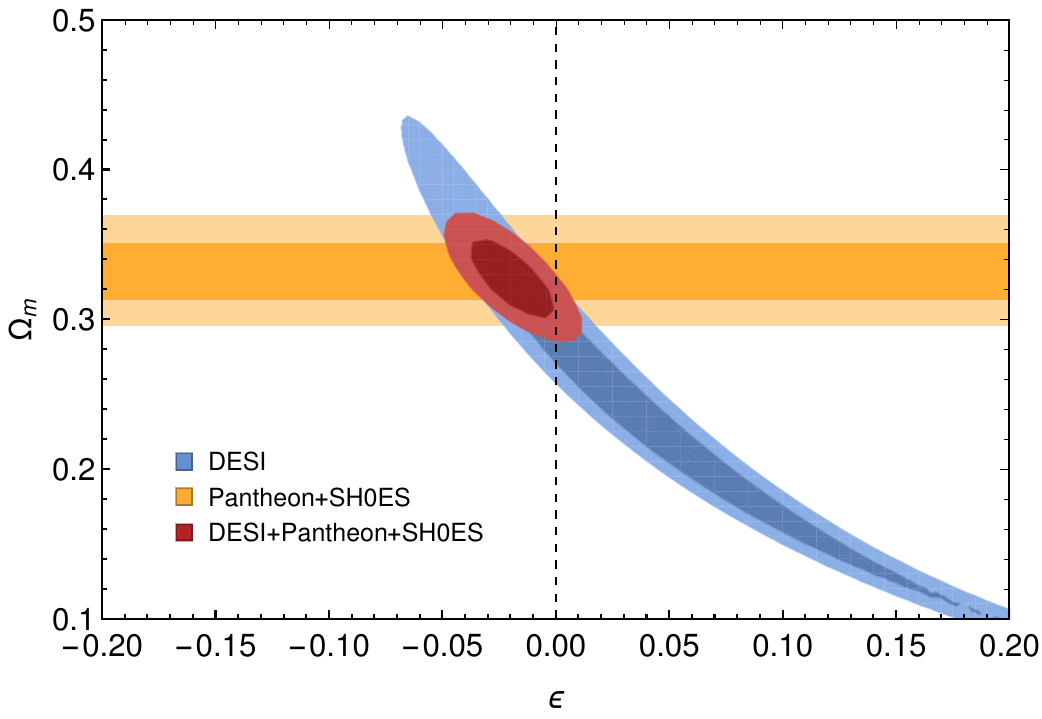}
	   \label{fig:Om-eps-dL-dim-cepheids}}
    \hspace{0.5cm}
    \subfloat[{Contour plot for Pantheon+SH0ES (yellow), DESI+BBN (blue) and their combination (red).}
    ]{
	   \includegraphics[width=7cm, clip]{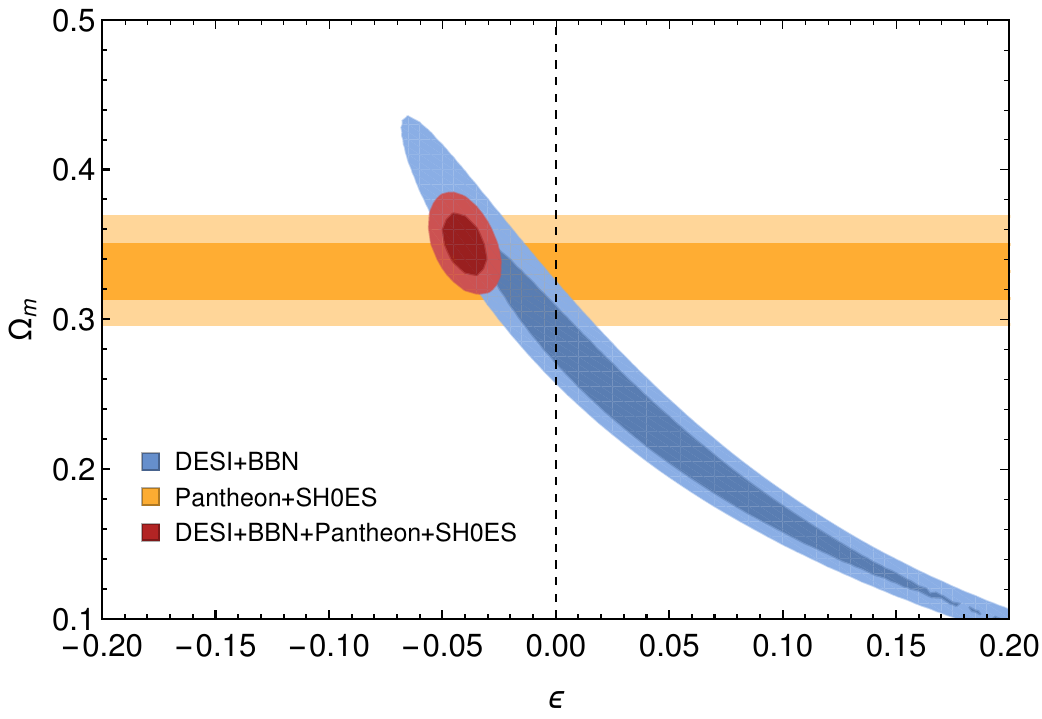}
	   \label{fig:Om-eps-dL-dim_tension}}
    \caption{Evaluation of the profile likelihood of $\Om$ and $\epsilon$ with Pantheon+ and DESI DR1 data for different cases.}
    \label{Pantheon-contours}
\end{figure}

We sample the parameters $\epsilon$ (defined in Eq. \ref{eta-z-epsilon}) and $\Om$ using a profile likelihood approach and sampling a grid since this does not require much computational time. The profile likelihood contour plot for the standard case is shown in Fig. \ref{fig:Om-eps-dL-dim}. The best-fit point is $\epsilon=-0.019$ with a distance from $\epsilon=0$ of $1.54\sigma$. 

Pantheon+ finds a value of $\Om=0.334\pm0.018$ \cite{Brout:2022vxf} and DESI DR1 computes a lower value of $\Om=0.295\pm0.0015$. Since this constitutes a discrepancy of $2.2\sigma$, it is not surprising that we also find an $\epsilon$ different form $0$ using these data sets. This is also reflected in the incompatibility of $\omega_m = \Omega_m h^2$ values between early and late-time measurements, see \cite{Blanchard:2022xkk}.

In a next step, we add data sets to our likelihoods that are in tension with respect to the Hubble parameter $H_0$ to reflect this tension in a violation of the DDR. As detailed in Sect. \ref{subsec:BBN}, for the second case, we add BBN information to the BAO data, which makes the likelihood sensitive to $H_0$. The result is plotted in Fig. \ref{fig:Om-eps-dL-dim-planck}. This is extremely similar to Fig. \ref{fig:Om-eps-dL-dim}, since the analytic marginalisation over the absolute SNe Ia magnitude $M_0$ can compensate for a value of $H_0$ that would not match the SNe Ia data in the calibrated case.

For the third case, we calibrate the Pantheon+ SNe Ia using Cepheids with the SH0ES data set, as was also done in the Pantheon+ analysis \cite{Brout:2022vxf} but we do not use BBN information. This is shown in Fig. \ref{fig:Om-eps-dL-dim-cepheids}. Here as well, there is not much visible change, compared to Fig. \ref{fig:Om-eps-dL-dim}. This is because the BAO likelihood is minimised over $r_{\rm d}\cdot h$, essentially leaving $H_0$ adjustable to the SH0ES value.

Lastly, we combine the last two cases and use both the Cepheid calibration and the BBN value of $\ob$. Now, both likelihoods are sensitive to $H_0$ but prefer a different value. This, in turn, creates a violation of the DDR which can be seen in Fig. \ref{fig:Om-eps-dL-dim_tension}. For the combined contour, the best-fit is at $\epsilon=-0.040$ and $6.2 \sigma$ away from 0. DESI+BBN finds $H_0=68.53\pm0.80\;\rm{km}\;\rm{s}^{-1}\;\rm{Mpc}^{-1}$ \cite{DESI:2024mwx}, whereas Pantheon+SH0ES lies at $H_0=73.5\pm1.1\;\rm{km}\;\rm{s}^{-1}\;\rm{Mpc}^{-1}$ \cite{Brout:2022vxf}. This constitutes a $4.5\sigma$ difference in $\Lambda$CDM. Together with the discrepancy in $\Om$, it makes sense that we find a value of $\epsilon$ that is far from $\epsilon=0$.

\begin{table}[h]
    \centering
        \begin{tabular}{ c | c | c | c | c}
            DESI+Pantheon+ & Uncalibrated & BBN & SH0ES & BBN+SH0ES \\
            \hline
            $\nu$ for $\Lambda$CDM & 1598 & 1599 & 1665& 1666 \\
            \hline
            $\nu$ for $\epsilon\neq 0$ & 1597 & 1598 & 1664 & 1665 \\
            \hline
            $\chi^2$ for $\Lambda$CDM & 1429.23 & 1429.23 & 1469.03 & 1507.68 \\
            \hline
            $\chi^2$ for $\epsilon\neq 0$ & 1426.83  & 1426.83 & 1466.61& 1470.77 \\
            \hline
            $\Delta\chi^2$ & -2.40 & -2.40 & -2.67& -36.72\\
            \hline
            $P(\nu,\chi^2)$ for $\Lambda$CDM & 0.99898 & 0.99904& 0.99979& 0.99762 \\
            \hline
            $P(\nu,\chi^2)$ for $\epsilon\neq 0$ & 0.99907& 0.99913 & 0.99981&0.99976 \\
            \hline
            $\Delta P(\nu,\chi^2)$ & 0.00009 & 0.00009 & 0.00002& 0.00214\\
            \hline
            Best-fit $\epsilon$ & -0.019 $\pm$ 0.012 & -0.019 $\pm$ 0.012 & -0.019 $\pm$ 0.012 & -0.040 $\pm$ 0.006
        \end{tabular}
        \caption{Comparison of the degrees of freedom $\nu$, the $\chi^2$ and $P(\nu,\chi^2)$ values for DESI DR1 and Pantheon+ data in the $\Lambda$CDM and the modified DDR model. The best-fit value for the $\epsilon$ parameter is also reported for each case.}.
        \label{chi2-pantheon}
\end{table}
\raggedbottom
Table \ref{chi2-pantheon} summarises the degrees of freedom $\nu$, the $\chi^2$ and $P(\nu,\chi^2)$ values (see Sect.\ref{subsec:goodness-of-fit}) for the four cases that we outlined. The $\Lambda$CDM value is compared to the value for the $\epsilon$-model and we report the best-fit value for the $\epsilon$-parameter in each case as well. We provide the results for BBN+SH0ES for $\Lambda$CDM to show that the fit is not as good compared to DDR, and to illustrate the inconsistency between the two calibrations within $\Lambda$CDM. However, we are aware that within $\Lambda$CDM, BBN and SH0ES should in principle not be combined given the Hubble tension.

As discussed in Sect. \ref{subsec:SN data}, in the uncalibrated case, all SNIa with $z<0.01$ are removed from the analysis, leaving 1590 out of 1701 objects. For the calibrated case using SH0ES, SNIa are removed that are below this threshold and not a calibrator. In that case, the sample has 1657 entries, since most low redshift SNIa are calibrators. Thus, the large difference for the $\chi^2$ values between dis- and enabling Cepheid calibration comes from the different amount of data points in the sample.

In the last case, using BBN and SH0ES, we find a $\chi^2 \approx 1508$ in $\Lambda$CDM and a much improved value of $\chi^2 \approx 1471$ when allowing for $\epsilon \neq 0$. The $P(\nu,\chi^2)$ values are all close to 100\%, since $\Lambda$CDM is already a very good fit for this many data points and only few parameters. However, the improvement when violating the DDR in the BBN+SH0ES case is still more than 20 times greater than in the other cases, indicating an alleviation of the Hubble tension.

\begin{figure}[h]
	\center
	\includegraphics[width=7cm]{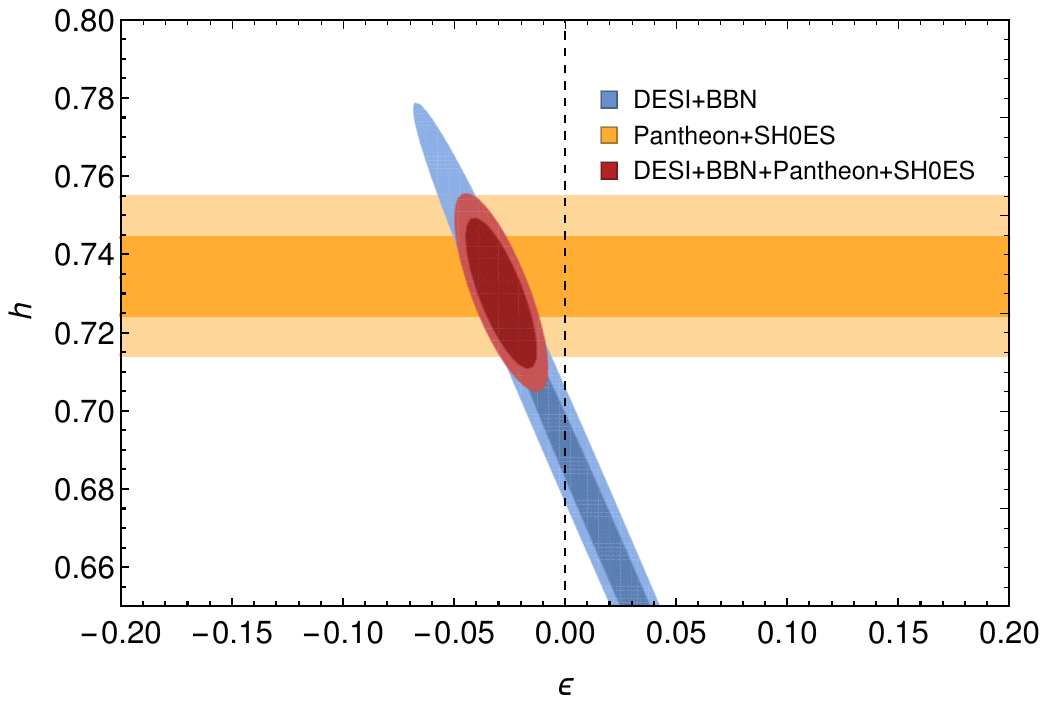}
	\caption{Contour plot of $h$ and $\epsilon$ for Pantheon+SH0ES (yellow), DESI+BBN (blue) and their combination (red).}
	\label{fig:Panth-DESI-tension-h}
\end{figure}

For this case, we also plot the contours of $\epsilon$ and the dimensionless Hubble parameter $h$ in Fig. \ref{fig:Panth-DESI-tension-h}. The contours show that the $\epsilon$-parameter is correlated with $h$ which allows our model to alleviate the Hubble tension.

\subsubsection*{Transverse BAO}

In the DESI BAO data release, per redshift bin, either both the comoving distance $d_{\rm M}$ and the Hubble distance $d_{\rm H}$ or only the volume-averaged distance $d_{\rm V}$ is reported. To test for example the conservation of photons, it is useful to incorporate all distance measurements into the analysis. This is common in the literature, see e.g. \cite{Euclid:2020ojp, Alfano:2025gie, Jesus:2024nrl}.
\begin{figure}[h]
	\center
	\includegraphics[width=7cm]{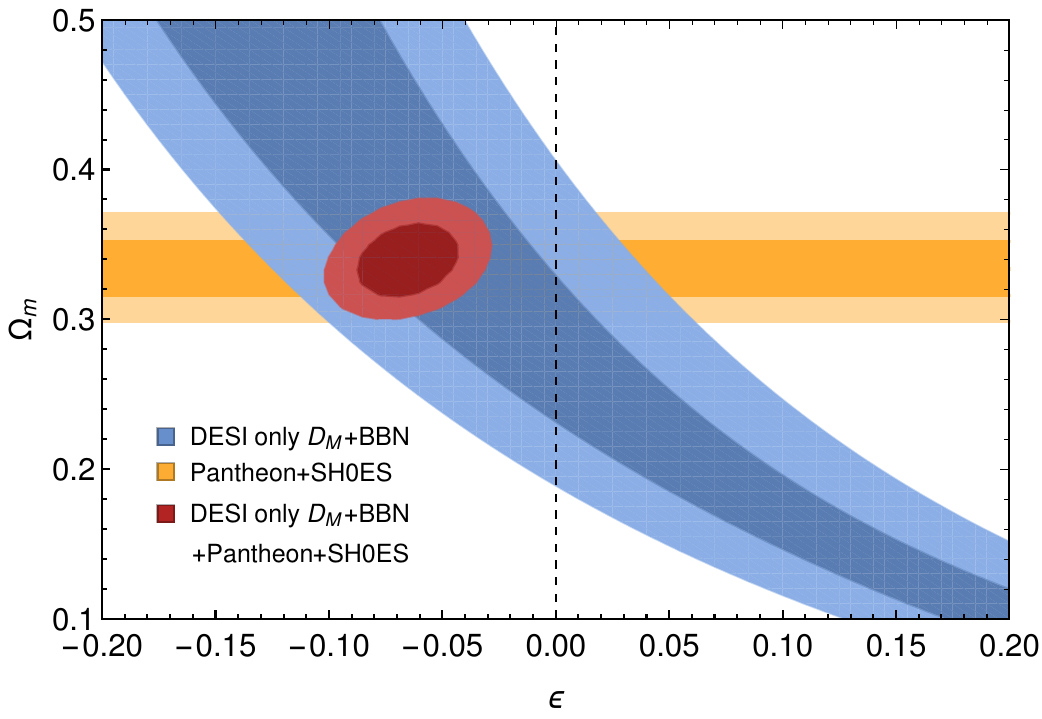}
	\caption{Contour plot of $\Om$ and $\epsilon$ for Pantheon+SH0ES (yellow), DESI+BBN (blue) and their combination (red) using only the comoving distance measurements $d_{\rm M}$ from DESI.}
	\label{fig:Panth-DESI-only-dm}
\end{figure}
The DDR technically only concerns the angular diameter distance which is transverse to the line-of-sight and not the Hubble distance or the volume-averaged distance which contains the Hubble distance. Thus, to be more conservative, it is possible to only use the data points of the transverse comoving distance $d_{\rm M}=(1+z)d_{\rm A}$. We recalculate the earlier case with BBN and Cepheid calibration (see Fig. \ref{fig:Om-eps-dL-dim_tension}) with only these values to show this difference in Fig. \ref{fig:Panth-DESI-only-dm}. The BAO contours are much larger in this case because only 5 instead of 12 data points from DESI are used. This also increases the combined contour. However, the preferred value is $\epsilon=-0.065$, which is lower than in Fig. \ref{fig:Om-eps-dL-dim_tension} and still $4.4\sigma$ apart from the DDR. We note that to focus on this aspect, it would help to add other BAO data sets to DESI. In the rest of this work, we will use all 12 DESI data points for the GA and defer this more conservative test to future work.

\subsubsection{DES-SN5YR with DESI DR1}
\label{sec:DES-SN5YRwithDESI}
\begin{figure}[h!]
    \centering
    \subfloat[{Contour plot for DES-SN5YR (green), DESI (blue) and their combination (brown).}
        ]{
        \includegraphics[width=7cm, clip]{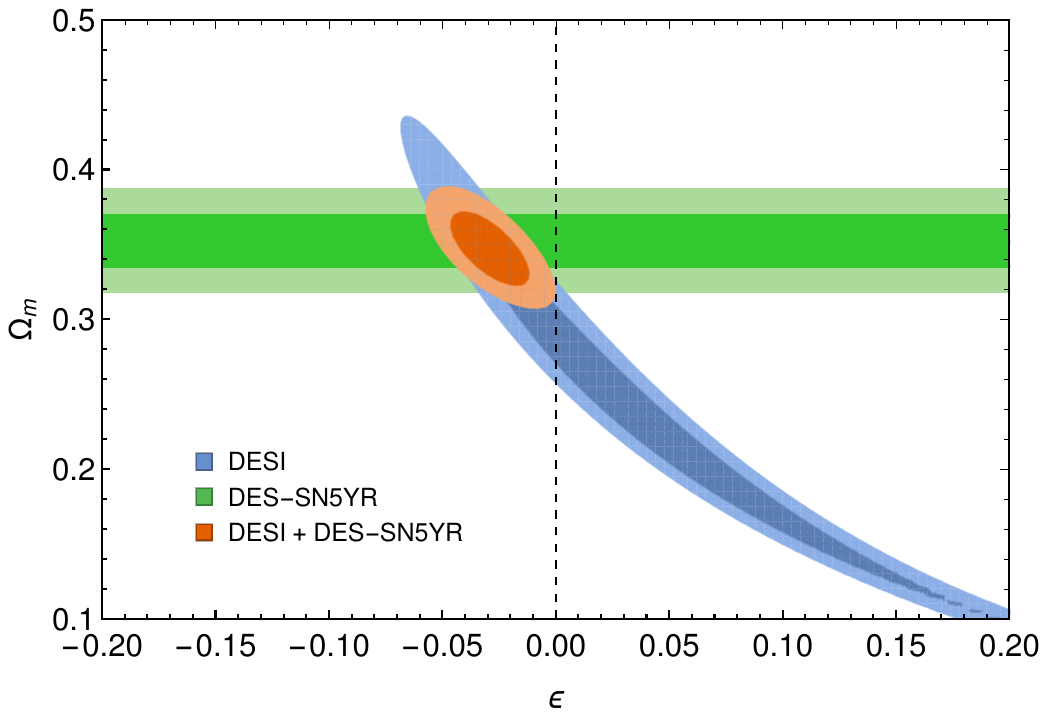}
        \label{fig:Om-eps-dL-des}}
    \hspace{0.5cm}
    \subfloat[{Contour plot for DES-SN5YR (green), DESI+BBN (blue) and their combination (brown).}
        ]{
        \includegraphics[width=7cm, clip]{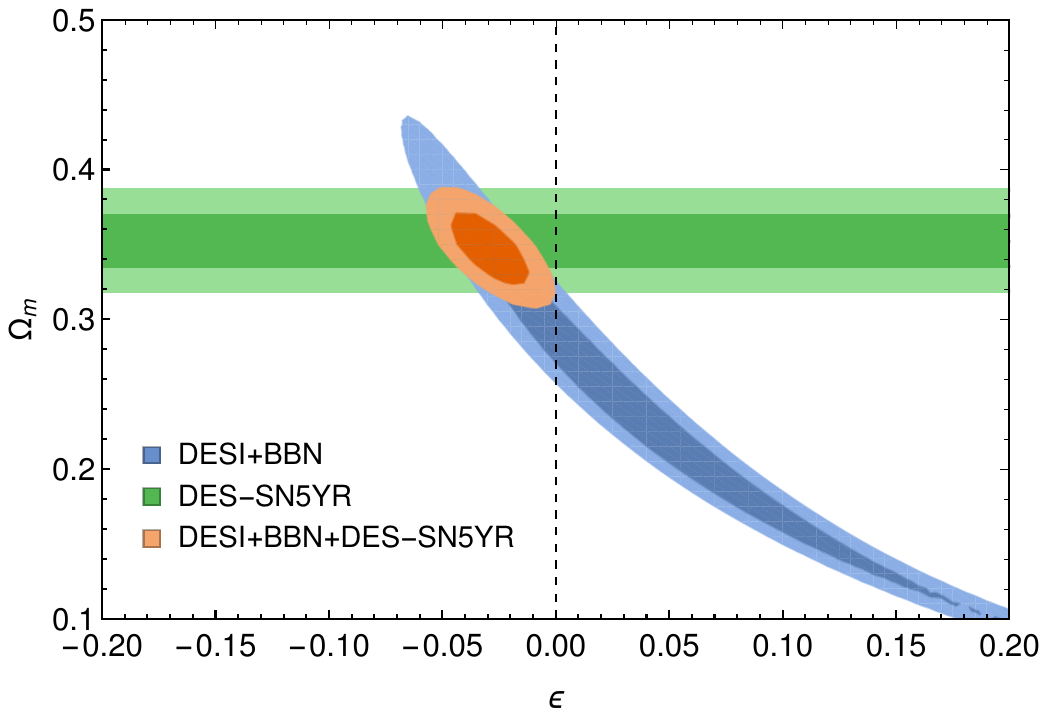}
        \label{fig:Om-eps-dL-des-planck}}
    \\
    \subfloat[{Contour plot for DES-SN5YR+SH0ES (green), DESI (blue) and their combination (brown).}
    ]{
	   \includegraphics[width=7cm, clip]{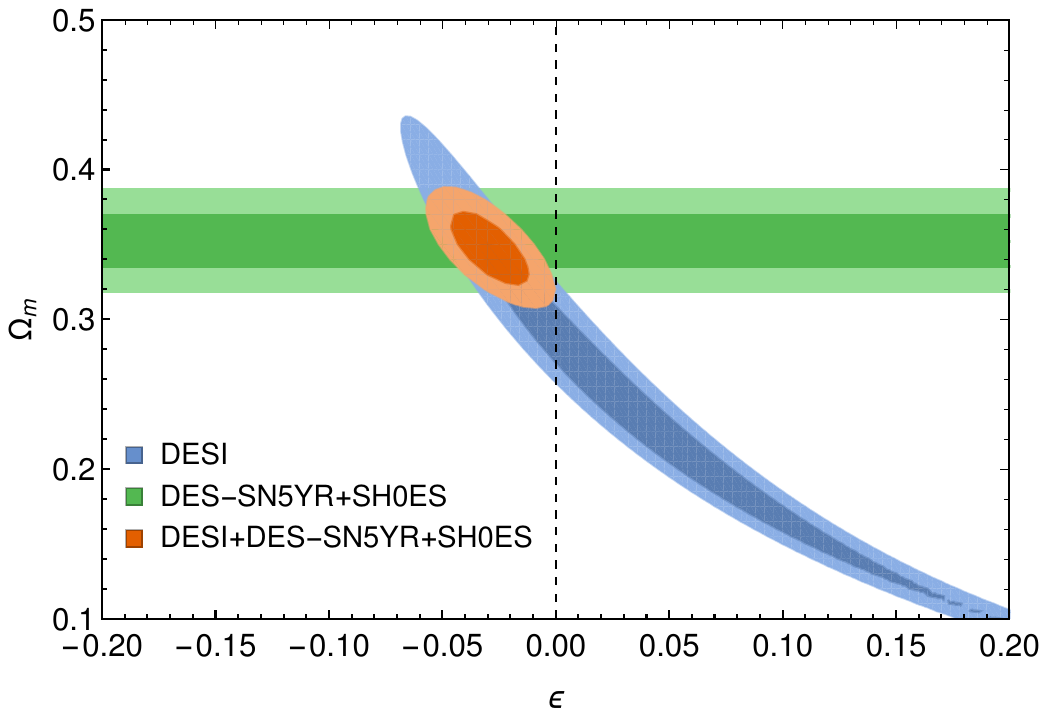}
	   \label{fig:Om-eps-dL-des-cepheids}}
    \hspace{0.5cm}
    \subfloat[{Contour plot for DES-SN5YR+SH0ES (green), DESI+BBN (blue) and their combination (brown).}
    ]{
	   \includegraphics[width=7cm, clip]{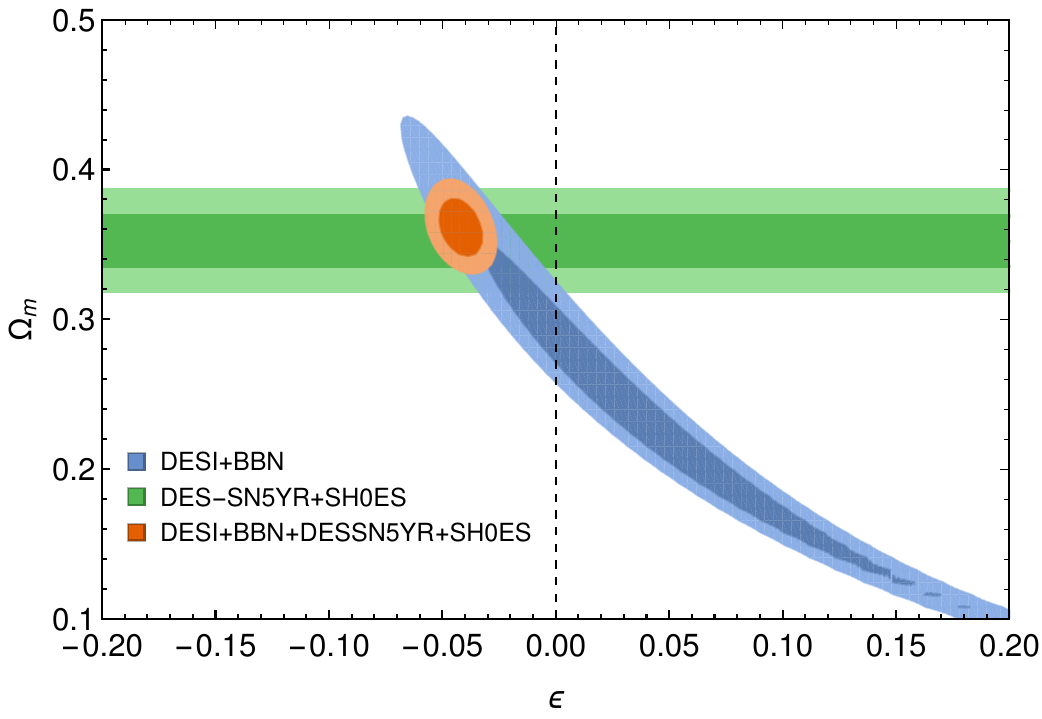}
	   \label{fig:Om-eps-dL-des-tension}}
    \caption{Evaluation of the profile likelihood of $\Om$ and $\epsilon$ with DES-SN5YR and DESI DR1 data for different cases.}
    \label{DES-contours}
\end{figure}

We also calculate the profile likelihood showing contour plots for $\Om$ and $\epsilon$, replacing the SNe Ia with the DES-SN5YR data set \cite{DES:2024jxu}. The results are plotted in Fig. \ref{DES-contours}. All four contours are similar to those for Pantheon+ in Fig. \ref{Pantheon-contours}. For the standard DES-SN5YR in Fig. \ref{fig:Om-eps-dL-des}, the best-fit value for $\epsilon$ is $2.5\sigma$ apart from $0$ compared to $2.2\sigma$ for Pantheon+. This is due to DES-SN5YR preferring a higher value of $\Om$, making it less compatible with DESI. When adding either only BBN (Fig. \ref{fig:Om-eps-dL-des-planck}) or only the $\chi^2$ from Eq. \ref{chi2-sh0es-h0} incorporating the $H_0$ value from Pantheon+SH0ES (Fig. \ref{fig:Om-eps-dL-des-cepheids}), the contours barely change with respect to the first case. In the last case, combining BBN and the SH0ES $\chi^2$, we find a strong violation of the DDR at $6.5\sigma$ compared to $6.2\sigma$ with Pantheon+. 

\begin{table}[h]
    \centering
        \begin{tabular}{ c | c | c | c | c}
            DESI+DES-SN5YR & Uncalibrated & BBN & SH0ES & BBN+SH0ES \\
            \hline
            $\nu$ for $\Lambda$CDM & 1837 & 1838& 1837& 1838\\
            \hline
            $\nu$ for $\epsilon\neq 0$ &1836 & 1837&1836 & 1837 \\
            \hline
            $\chi^2$ for $\Lambda$CDM & 1670.83 & 1670.83 & 1661.61 & 1698.18 \\
            \hline
            $\chi^2$ for $\epsilon\neq 0$ & 1664.51 &1664.51& 1655.36 & 1655.49 \\
            \hline
            $\Delta\chi^2$ & -6.32 & -6.32 & -6.25& -40.99\\
            \hline
            $P(\nu,\chi^2)$ for $\Lambda$CDM & 0.99758& 0.99771& 0.99856& 0.99077 \\
            \hline
            $P(\nu,\chi^2)$ for $\epsilon\neq 0$ &0.99820& 0.99830& 0.99894&0.99889 \\
            \hline
            $\Delta P(\nu,\chi^2)$ & 0.00062 & 0.00059& 0.00038& 0.00812\\
            \hline
            Best-fit $\epsilon$ & -0.029 $\pm$ 0.011 & -0.029 $\pm$ 0.011 & -0.029 $\pm$ 0.011 & -0.042 $\pm$ 0.006
        \end{tabular}
        \caption{Comparison of the degrees of freedom $\nu$, the $\chi^2$ and $P(\nu,\chi^2)$ values for DESI DR1 and DES-SN5YR data in the $\Lambda$CDM and the modified DDR model. The best-fit value for the $\epsilon$ parameter is also reported for each case.}
        \label{chi2-des}
\end{table}

Table \ref{chi2-des} summarises the degrees of freedom $\nu$, the $\chi^2$ and the $P(\nu,\chi^2)$ values (see Sect.\ref{subsec:goodness-of-fit}) for the four cases that we outlined for the DES-SN5YR data set. The $\Lambda$CDM value is compared to the value for the $\epsilon$-model and we also show the best-fit value for the $\epsilon$-parameter in each case. In the case of using BBN and SH0ES, we find a $\chi^2 \approx 1698$ in $\Lambda$CDM and a much smaller value of $\chi^2 \approx 1655$ when allowing for $\epsilon \neq 0$.

As discussed in the \ref{subsubsec:param-pantheon}, the $P(\nu, \chi^2)$ values are all close to 1, but we still see a much greater difference between $\Lambda$CDM and $\epsilon \neq 0$ in the last case using BBN+SH0ES.

\subsection{GA Best-Fit Functions}
We test five different cases with the GA, four using the Pantheon+ data set combined with SH0ES corresponding to the parametrised approach from Fig. \ref{Pantheon-contours} and one using the DES-SN5YR data set for the SNe Ia corresponding to Fig. \ref{fig:Om-eps-dL-des}. In all cases, we use DESI data for the BAO.

\subsubsection{Pantheon+ with DESI DR1}

\begin{figure}[h]
    \centering
    \subfloat[{GA Deviation from $\Lambda$CDM in percent for the angular diameter distance (red), the luminosity distance (green) and the Hubble parameter (purple).}
        ]{
        \includegraphics[width=7cm, clip]{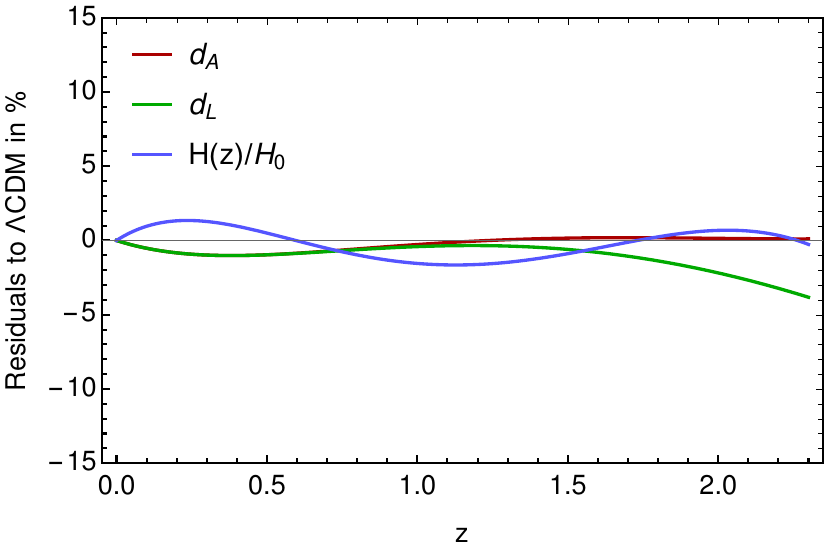}
        \label{fig:residuals-panth}}
    \hspace{0.5cm}
    \subfloat[{Deviation from the standard DDR $\eta(z)$ for the GA (in solid red) with the associated errors (shaded in orange) compared to $\Lambda$CDM (in dashed black), i.e. no deviation.}
        ]{
        \includegraphics[width=7cm, clip]{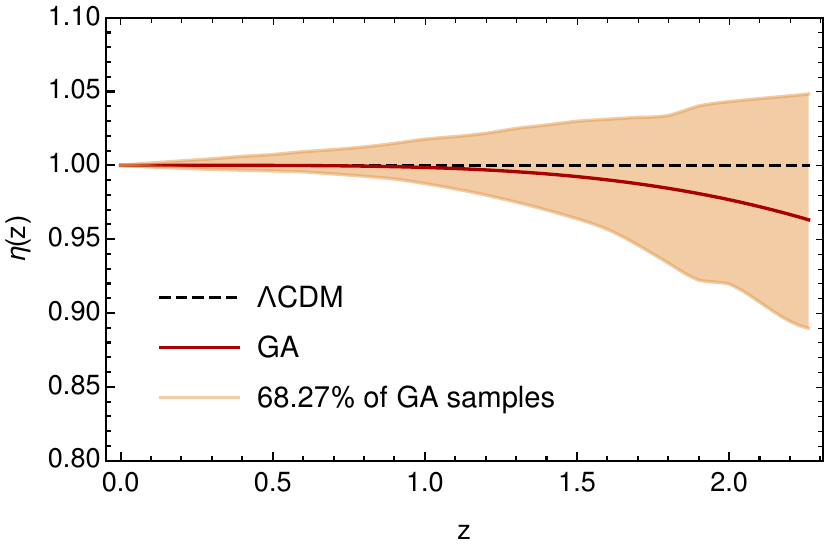}
        \label{fig:eta_of_z_errors}}
    \caption{Results of the GA using Pantheon+ and DESI DR1 data.}
    \label{fig:Panth-results}
\end{figure}

When applying the GA to the uncalibrated SNe Ia and BAO data, we obtain the results of the GA for the angular diameter distance, the luminosity distance, and the Hubble parameter. We show the difference of each of these functions with respect to the best-fit $\Lambda$CDM in percent in Fig. \ref{fig:residuals-panth}. For the angular diameter distance, no discrepancy to $\Lambda$CDM can be seen. The luminosity distance deviates slightly below $\Lambda$CDM as is visible in the DDR deviation as well. The Hubble parameter also shows no deviation from the standard model.
The GA result for the DDR violation $\eta(z)=d_{\rm L}(z)/[d_{\rm A}(z)(1+z)^2]$ is shown in Fig. \ref{fig:eta_of_z_errors} where $\eta(z)=1$ corresponds to no violation. This is consistent with the deviation of $\epsilon=-0.0192$ we find in the parametrised approach in Sect. \ref{subsec:parametrised-results}. However, here in the unparametrised approach this change constitutes less than $0.5\sigma$ using our sampled error.

\paragraph{BBN and \textit{Planck} Values}\label{Pantheon-Planck}

\begin{figure}[h]
    \centering
    \subfloat[{GA Deviation from $\Lambda$CDM in percent for the angular diameter distance (red), the luminosity distance (green) and the Hubble parameter (purple).}
        ]{
        \includegraphics[width=7cm, clip]{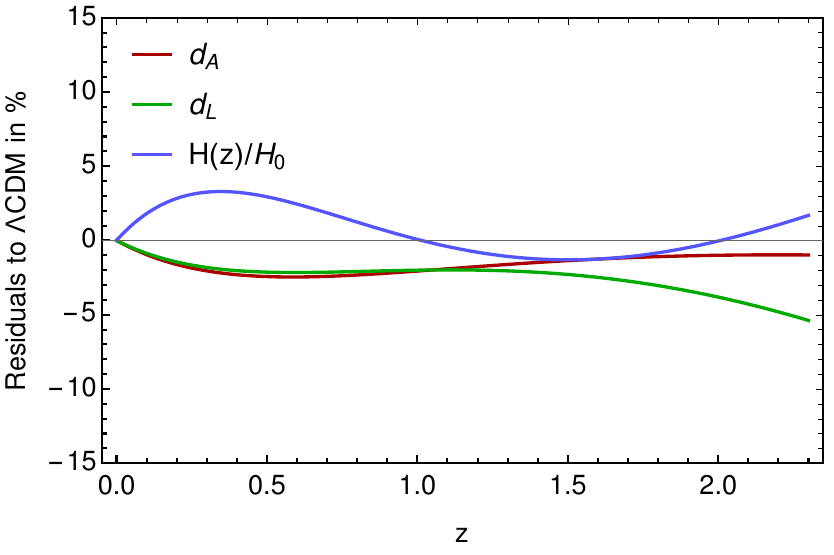}
        \label{fig:residuals-panth-planck}}
    \hspace{0.5cm}
    \subfloat[{Deviation from the standard DDR $\eta(z)$ for the GA (in solid red) with the associated errors (shaded in orange) compared to $\Lambda$CDM (in dashed black), i.e. no deviation.}
        ]{
        \includegraphics[width=7cm, clip]{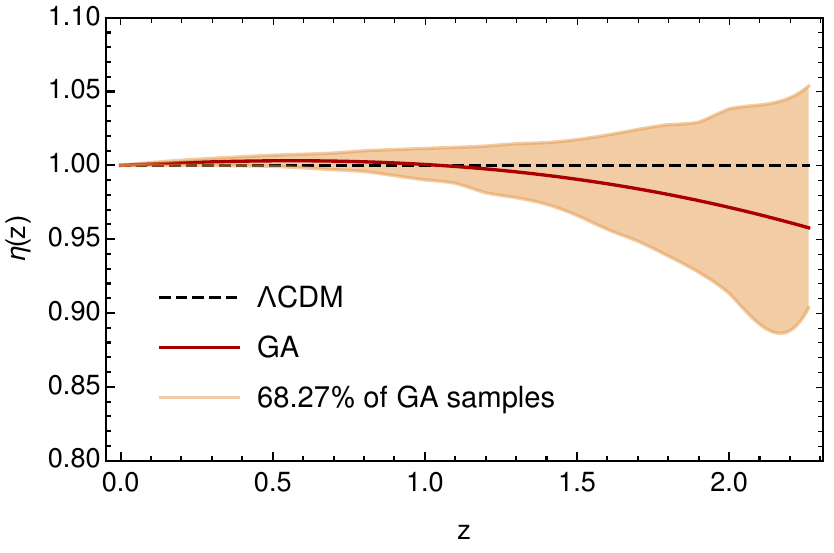}
        \label{fig:eta_of_z_pantheon_planck}}
    \caption{Results of the GA using Pantheon+ and DESI DR1 with BBN and \textit{Planck} values.}
    \label{fig:Panth-results-planck}
\end{figure}

As described in Sect. \ref{sec:GA}, we show the deviation from the DDR using uncalibrated SNe Ia and setting $\ob$ to the BBN value and $\om$ to the \textit{Planck} 2018 value in Fig. \ref{fig:Panth-results-planck}. The residuals with respect to $\Lambda$CDM are slightly larger than in the standard case but still below 5\%, see Fig. \ref{fig:residuals-panth-planck}. The evolution of $\eta(z)$ looks very similar to \ref{fig:eta_of_z_errors} from the standard analysis. This was expected since DESI DR1 is mostly in agreement with \textit{Planck} 2018 and the absolute SNe Ia magnitude is marginalised over analytically, which essentially leaves the amplitude of the luminosity distance as a free parameter. This can then be adjusted to fit together with the amplitude of the BAO data.


\begin{figure}[h]
    \centering
    \subfloat[{GA Deviation from $\Lambda$CDM in percent for the angular diameter distance (red), the luminosity distance (green) and the Hubble parameter (purple).}
        ]{
        \includegraphics[width=7cm, clip]{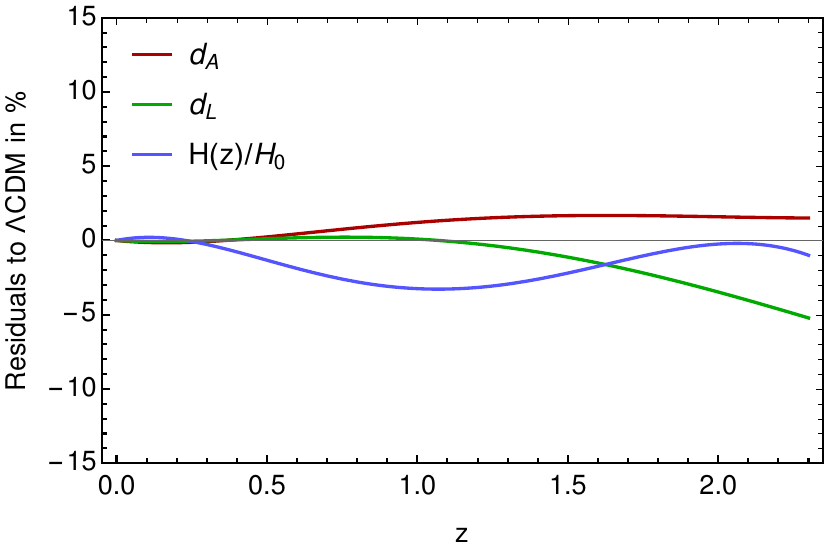}
        \label{fig:residuals-panth-cepheid}}
    \hspace{0.5cm}
    \subfloat[{Deviation from the standard DDR $\eta(z)$ for the GA (in solid red) with the associated errors (shaded in orange) compared to $\Lambda$CDM (in dashed black), i.e. no deviation.}
        ]{
        \includegraphics[width=7cm, clip]{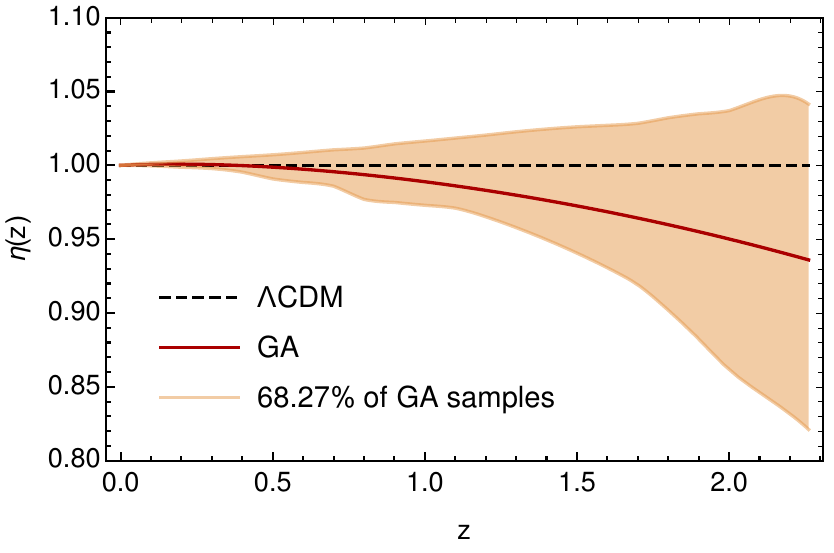}
        \label{fig:eta_of_z_pantheon}}
    \caption{Results of the GA using using Pantheon+SH0ES and DESI DR1.}
    \label{fig:Panth-results-cepheid}
\end{figure}

\paragraph{Cepheid Calibration}
When enabling the Cepheid calibration (see Sect. \ref{subsubsec:Cali}), the GA finds the results shown in Fig. \ref{fig:Panth-results-cepheid}. Here, the deviations from $\Lambda$CDM are larger than in Fig. \ref{fig:Panth-results} and Fig. \ref{fig:Panth-results-planck} but they remain below 5\% for the residuals in Fig. \ref{fig:residuals-panth-cepheid} and the error on $\eta(z)$ is smaller than 1$\sigma$.

\paragraph{Cepheid Calibration and \textit{Planck} Values}
Here, we test the GA for the case in which the SNe Ia data reflect the Pantheon+ with SH0ES $H_0$ value of $H_0 = 73.5\pm 1.1\, \kmsMpc $ in $\Lambda$CDM \cite{Brout:2022vxf}. At the same time the BAO reflect the \textit{Planck} 2018 \cite{Planck:2018vyg}  values of $\om$ and $\ob$. Together with the DESI BAO this yields $H_0=68.53\pm 0.80 \, \kmsMpc$ in $\Lambda$CDM \cite{DESI:2024mwx}.

\begin{figure}[h]
    \centering
    \subfloat[{GA Deviation from $\Lambda$CDM in percent for the angular diameter distance (red), the luminosity distance (green) and the Hubble parameter (purple).}
        ]{
        \includegraphics[width=7cm, clip]{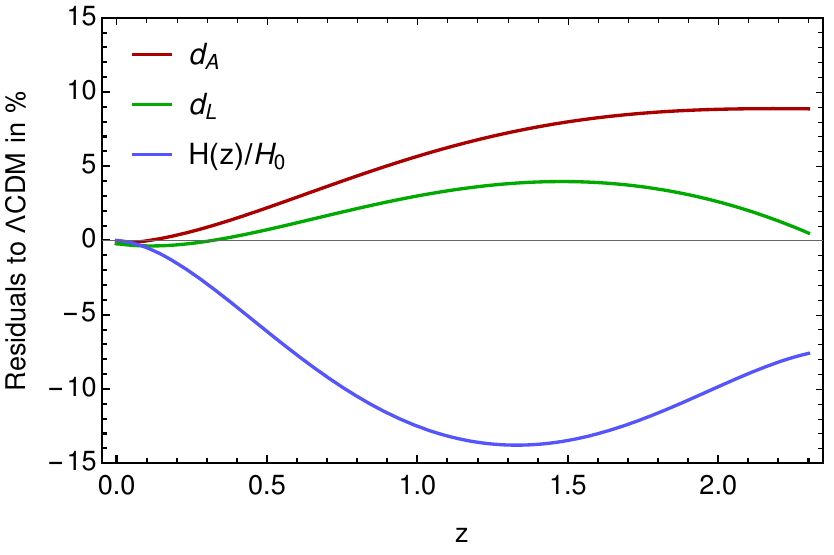}
        \label{fig:residuals-panth-cepheid-planck}}
    \hspace{0.5cm}
    \subfloat[{Deviation from the standard DDR $\eta(z)$ for the GA (in solid red) with the associated errors (shaded in orange) compared to $\Lambda$CDM (in dashed black), i.e. no deviation.}
        ]{
        \includegraphics[width=7cm, clip]{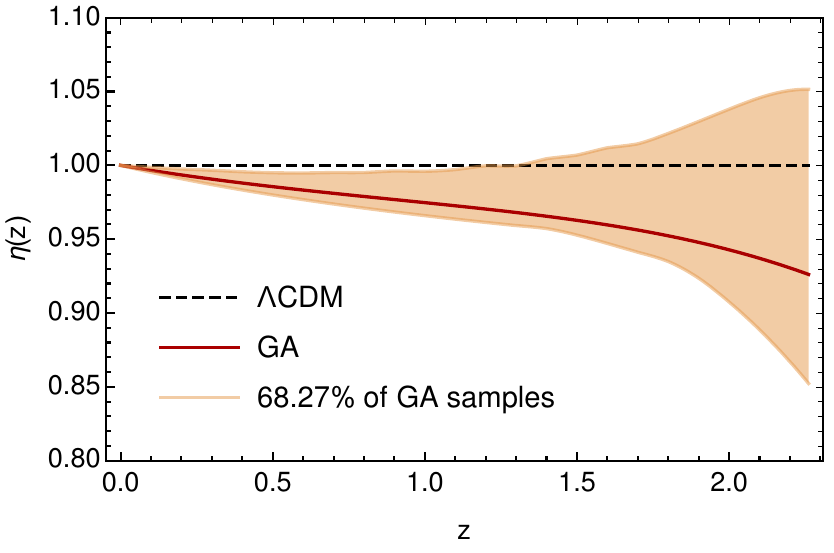}
        \label{fig:eta_of_z_tension_errors}}
    \caption{Results of the GA using using Pantheon+SH0ES and DESI DR1 with BBN and \textit{Planck} values.}
    \label{fig:Panth-results-planck-cepheid}
\end{figure}

Fig. \ref{fig:residuals-panth-cepheid-planck} again shows the results of the GA for the angular diameter distance, the luminosity distance, and the Hubble parameter.  The angular diameter distance found by the GA is much higher than in $\Lambda$CDM. In contrast to the standard case, the $H(z)$ predicted by the GA is much lower than in the standard model at higher redshifts. This is expected since at lower $z$, the SNe Ia data with the higher $H_0$ dominate while for higher redshifts BAO+BBN data dominate which prefer a lower $H_0$. The deviation from the DDR can be seen in Fig. \ref{fig:eta_of_z_tension_errors}. This shows a much stronger violation than in the standard case, see Fig. \ref{fig:eta_of_z_errors}. This is consistent with our results in the parametrised approach that found $\epsilon=-0.040$. These results reflect the Hubble tension present in the combined data sets very well. However, with this model-independent approach the error remains at $ 1\sigma$.

\subsubsection{DES-SN5YR with DESI DR1}


\begin{figure}
    \centering
    \subfloat[{GA Deviation from $\Lambda$CDM in percent for the angular diameter distance (red), the luminosity distance (green) and the Hubble parameter (purple).}
        ]{
        \includegraphics[width=7cm, clip]{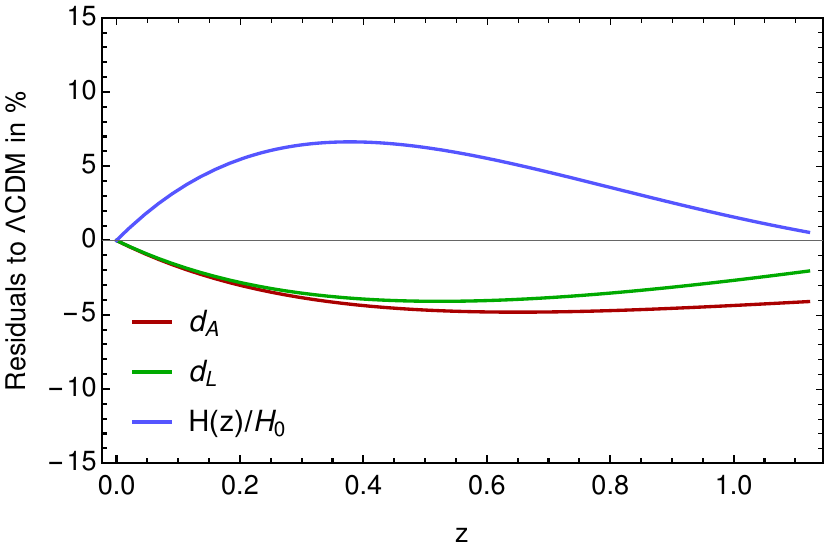}
        \label{fig:residuals-des}}
    \hspace{0.5cm}
    \subfloat[{Deviation from the standard DDR $\eta(z)$ for the GA (in solid red) with the associated errors (shaded in orange) compared to $\Lambda$CDM (in dashed black), i.e. no deviation.}
        ]{
        \includegraphics[width=7cm, clip]{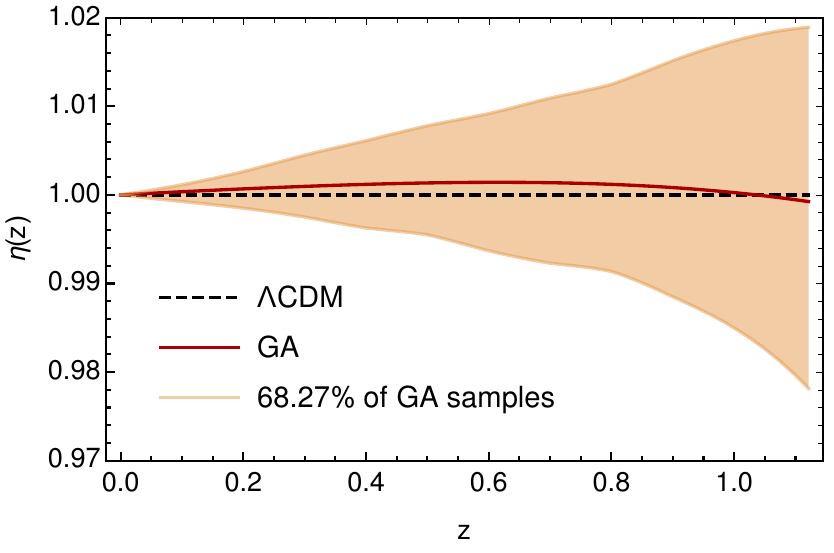}
        \label{fig:eta_of_z_errors-des}}
    \caption{Results of the GA using using DES-SN5YR and DESI DR1.}
    \label{fig:DES-results}
\end{figure}

We also implement the DES-SNYR5 \cite{DES:2024jxu} likelihood and use the GA on this data set in combination with DESI DR1. In Fig. \ref{fig:DES-results}, the GA result for both distances is shown. Since DES-SN5YR covers a redshift range up to 1.13, which is roughly half of the Pantheon+ range, we shrink the $\eta$ axis range to $0.95$--$1.05$. There is no discrepancy to be seen with respect to $\Lambda$CDM. This is mirrored in Fig. \ref{fig:eta_of_z_errors-des} which also shows the sampled error for $\eta(z)$. The error for the GA lies slightly below $2\%$ in this case which is much lower than for the Pantheon+ case. For this SNe Ia data set, we only consider this standard case since the results in the parametrised approach are very similar, and the GA including the error calculation is computation intensive.
\section{Conclusion}\label{sec:conclusion}

This work tests the cosmic DDR using new data sets and provides a new way of looking at the Hubble tension. We use a parametrised approach and a model-agnostic GA to constrain the violation of the cosmic DDR for the Pantheon+ or the DES-SN5YR SNe Ia in combination with the DESI DR1 BAO data set. For the parametrised, model-dependent approach, we calculate the profile likelihood to find contours in the $\Om$--$\epsilon$ plane, where $\epsilon$ quantifies deviations from the DDR given by Eq. \ref{eta-z-epsilon}. We find that the best-fit point is $\epsilon=-0.019$ and $1.5\sigma$ apart from 0. The reported $\Om$ values from both surveys are over $2\sigma$ apart, which explains this value. 

When calibrating the SNe Ia with Cepheids using SH0ES and fixing the value of $\ob$ to the value from BBN, we find the best-fit for $\epsilon$ at $-0.040$ which is $6.2\sigma$ from $0$. This is due to the fact that both likelihoods are sensitive to the Hubble constant in this combination and the discrepancy of the two data sets lies at $4.5\sigma$ in $\Lambda$CDM. With the DES-SN5YR data set, we find very similar results when combining with BBN and SH0ES, yielding a $6.5\sigma$ discrepancy.

Then, we employ GA to test this in a model-independent way. For the standard, uncalibrated case, we find a very slight difference from $\eta(z)=1$ which indicates no DDR deviations. To calculate the error, we run the GA 100 times for each case and select 68.27\% of functions that are closest to our best-fit for an estimation of the $1\sigma$ error, analogously to the Gaussian error. In the standard analysis, this error is much larger than the DDR discrepancy.

In a next step, the BBN value for $\ob$ and \textit{Planck} values for $\om$ are assumed in the sound horizon at drag redshift when computing the DESI BAO likelihood. This does not significantly change the result for $\eta(z)$. The third case considers SH0ES Cepheid-calibrated SNe Ia with the standard DESI likelihood. There, we find a slightly stronger deviation that starts at much earlier redshifts.

When combining the last two cases, i.e. BBN and \textit{Planck} values for BAO and Pantheon+SH0ES, we run the analysis using the dimension-full luminosity and angular diameter distances. In this case, we see a much stronger violation of the DDR which is also reflected in the $H(z)$ calculated from the GA result. This is confirmed when looking at the sampled error. At redshifts below 1.4, the DDR violation is close to our sampled error corresponding to $1 \sigma$.

We also consider DES-SN5YR using GA. There, the maximum redshift lies at $1.13$, which is less than half of Pantheon+. In that case, there is no preference for a DDR violation. Since the parametrised results for both data sets are extremely similar, we do not test all cases with the GA since they are computationally expensive.

In conclusion, we show that the DDR is slightly violated in the uncalibrated parametrised approach and more strongly violated when adding BBN and the SH0ES calibration, reflecting the Hubble tension. This holds for both considered SNe Ia data sets. When using a model-independent approach, the DDR is confirmed within 1$\sigma$. This emphasises the importance of model-independent approaches for possible DDR violations.

\section*{Acknowledgements}
We would like to thank Matteo Martinelli for useful discussions and Sefa Pamuk for proofreading and providing valuable comments.
FK thanks the Instituto de Física Teórica (IFT) UAM-CSIC in Madrid, Spain, for their hospitality during the stay. This study has been partially supported through the grant EUR TESS N°ANR-18-EURE-0018 in the framework of the Programme des Investissements d'Avenir.
SN acknowledges support from the research project PID2021-123012NB-C43 and the Spanish Research Agency (Agencia Estatal de Investigaci\'on) through the Grant IFT Centro de Excelencia Severo Ochoa No CEX2020-001007-S, funded by MCIN/AEI/10.13039/50110 0011033. 
IT has been supported by the Ramon y Cajal fellowship (RYC2023-045531-I) funded by the State Research Agency of the Spanish Ministerio de Ciencia, Innovaci\'on y Universidades, MICIU/AEI/10.13039/501100011033, and Social European Funds plus (FSE+). IT also acknowledges support from the same ministry, via projects PID2019-11317GB, PID2022-141079NB, PID2022-138896NB; the European Research Executive Agency HORIZON-MSCA-2021-SE-01 Research and Innovation programme under the Marie Sk\l odowska-Curie grant agreement number 101086388 (LACEGAL) and the programme Unidad de Excelencia Mar\'{\i}a de Maeztu, project CEX2020-001058-M.

\bibliography{Distance_Duality_GA_inspire.bib}

@article{Nesseris:2012tt,
    author = "Nesseris, Savvas and Garcia-Bellido, Juan",
    title = "{A new perspective on Dark Energy modeling via Genetic Algorithms}",
    eprint = "1205.0364",
    archivePrefix = "arXiv",
    primaryClass = "astro-ph.CO",
    reportNumber = "IFT-UAM-CSIC-12-42",
    doi = "10.1088/1475-7516/2012/11/033",
    journal = "JCAP",
    volume = "11",
    pages = "033",
    year = "2012"
}

@article{Euclid:2020ojp,
    author = "Martinelli, M. and others",
    collaboration = "Euclid",
    title = "{Euclid: Forecast constraints on the cosmic distance duality relation with complementary external probes}",
    eprint = "2007.16153",
    archivePrefix = "arXiv",
    primaryClass = "astro-ph.CO",
    reportNumber = "IFT-UAM/CSIC-20-117",
    doi = "10.1051/0004-6361/202039078",
    journal = "Astron. Astrophys.",
    volume = "644",
    pages = "A80",
    year = "2020"
}

@article{Brout:2022vxf,
    author = "Brout, Dillon and others",
    title = "{The Pantheon+ Analysis: Cosmological Constraints}",
    eprint = "2202.04077",
    archivePrefix = "arXiv",
    primaryClass = "astro-ph.CO",
    doi = "10.3847/1538-4357/ac8e04",
    journal = "Astrophys. J.",
    volume = "938",
    number = "2",
    pages = "110",
    year = "2022"
}

@article{SNLS:2011lii,
    author = "Conley, A. and others",
    collaboration = "SNLS",
    title = "{Supernova Constraints and Systematic Uncertainties from the First 3 Years of the Supernova Legacy Survey}",
    eprint = "1104.1443",
    archivePrefix = "arXiv",
    primaryClass = "astro-ph.CO",
    doi = "10.1088/0067-0049/192/1/1",
    journal = "Astrophys. J. Suppl.",
    volume = "192",
    pages = "1",
    year = "2011"
}

@article{Scolnic:2021amr,
    author = "Scolnic, Dan and others",
    title = "{The Pantheon+ Analysis: The Full Data Set and Light-curve Release}",
    eprint = "2112.03863",
    archivePrefix = "arXiv",
    primaryClass = "astro-ph.CO",
    doi = "10.3847/1538-4357/ac8b7a",
    journal = "Astrophys. J.",
    volume = "938",
    number = "2",
    pages = "113",
    year = "2022"
}

@article{Euclid:2021frk,
    author = "Nesseris, S. and others",
    collaboration = "Euclid",
    title = "{Euclid: Forecast constraints on consistency tests of the \ensuremath{\Lambda}CDM model}",
    eprint = "2110.11421",
    archivePrefix = "arXiv",
    primaryClass = "astro-ph.CO",
    reportNumber = "IFT-UAM/CSIC-21-117",
    doi = "10.1051/0004-6361/202142503",
    journal = "Astron. Astrophys.",
    volume = "660",
    pages = "A67",
    year = "2022"
}

@article{Jesus:2024nrl,
    author = "Jesus, Jos\'e F. and Gomes, Mikael J. S. and Holanda, Rodrigo F. L. and Nunes, Rafael C.",
    title = "{High-redshift cosmography with a possible cosmic distance duality relation violation}",
    eprint = "2408.13390",
    archivePrefix = "arXiv",
    primaryClass = "astro-ph.CO",
    doi = "10.1088/1475-7516/2025/01/088",
    journal = "JCAP",
    volume = "01",
    pages = "088",
    year = "2025"
}

@article{Wang:2025gus,
    author = "Wang, Qiumin and Cao, Shuo and Jiang, Jianyong and Zhang, Kaituo and Jiang, Xinyue and Liu, Tonghua and Mu, Chengsheng and Cheng, Dadian",
    title = "{New tests of cosmic distance duality relation with DESI 2024 BAO observations}",
    eprint = "2506.12759",
    archivePrefix = "arXiv",
    primaryClass = "astro-ph.CO",
    month = "6",
    year = "2025"
}

@article{Teixeira:2025czm,
    author = "Teixeira, Elsa M. and Giar\`e, William and Hogg, Natalie B. and Montandon, Thomas and Poudou, Ad\`ele and Poulin, Vivian",
    title = "{Implications of distance duality violation for the $H_0$ tension and evolving dark energy}",
    eprint = "2504.10464",
    archivePrefix = "arXiv",
    primaryClass = "astro-ph.CO",
    month = "4",
    year = "2025"
}

@article{Tripp:1997wt,
    author = "Tripp, Robert",
    title = "{A Two-parameter luminosity correction for type Ia supernovae}",
    reportNumber = "LBL-40857, LBNL-40857",
    journal = "Astron. Astrophys.",
    volume = "331",
    pages = "815--820",
    year = "1998"
}

@article{Avgoustidis:2010ju,
    author = "Avgoustidis, Anastasios and Burrage, Clare and Redondo, Javier and Verde, Licia and Jimenez, Raul",
    title = "{Constraints on cosmic opacity and beyond the standard model physics from cosmological distance measurements}",
    eprint = "1004.2053",
    archivePrefix = "arXiv",
    primaryClass = "astro-ph.CO",
    reportNumber = "DAMTP-2010-29, DESY-10-049, MPP-2010-43, ICCUB-10-026",
    doi = "10.1088/1475-7516/2010/10/024",
    journal = "JCAP",
    volume = "10",
    pages = "024",
    year = "2010"
}

@article{Mukherjee:2024ryz,
    author = "Mukherjee, Purba and Sen, Anjan Ananda",
    title = "{Model-independent cosmological inference post DESI DR1 BAO measurements}",
    eprint = "2405.19178",
    archivePrefix = "arXiv",
    primaryClass = "astro-ph.CO",
    doi = "10.1103/PhysRevD.110.123502",
    journal = "Phys. Rev. D",
    volume = "110",
    number = "12",
    pages = "123502",
    year = "2024"
}

@article{Raffelt:1987im,
    author = "Raffelt, Georg and Stodolsky, Leo",
    title = "{Mixing of the Photon with Low Mass Particles}",
    reportNumber = "MPI-PAE/PTh-54/87",
    doi = "10.1103/PhysRevD.37.1237",
    journal = "Phys. Rev. D",
    volume = "37",
    pages = "1237",
    year = "1988"
}

@article{Pan-STARRS1:2017jku,
    author = "Scolnic, D. M. and others",
    collaboration = "Pan-STARRS1",
    title = "{The Complete Light-curve Sample of Spectroscopically Confirmed SNe Ia from Pan-STARRS1 and Cosmological Constraints from the Combined Pantheon Sample}",
    eprint = "1710.00845",
    archivePrefix = "arXiv",
    primaryClass = "astro-ph.CO",
    doi = "10.3847/1538-4357/aab9bb",
    journal = "Astrophys. J.",
    volume = "859",
    number = "2",
    pages = "101",
    year = "2018"
}

@article{Schoneberg:2024ifp,
    author = {Sch\"oneberg, Nils},
    title = "{The 2024 BBN baryon abundance update}",
    eprint = "2401.15054",
    archivePrefix = "arXiv",
    primaryClass = "astro-ph.CO",
    doi = "10.1088/1475-7516/2024/06/006",
    journal = "JCAP",
    volume = "06",
    pages = "006",
    year = "2024"
}

@article{Yang:2025qdg,
    author = "Yang, Fan and Fu, Xiangyun and Xu, Bing and Zhang, Kaituo and Huang, Yang and Yang, Ying",
    title = "{Testing the cosmic distance duality relation using Type Ia supernovae and BAO observations}",
    eprint = "2502.05417",
    archivePrefix = "arXiv",
    primaryClass = "astro-ph.CO",
    doi = "10.1140/epjc/s10052-025-13892-w",
    journal = "Eur. Phys. J. C",
    volume = "85",
    number = "2",
    pages = "186",
    year = "2025"
}

@book{Wolschin:2010skf,
    editor = "Wolschin, Georg",
    title = "{Lectures on cosmology}: {accelerated expansion of the Universe}",
    doi = "10.1007/978-3-642-10598-2",
    publisher = "Springer",
    address = "Berlin",
    year = "2010"
}

@article{Alfano:2025gie,
    author = "Alfano, Anna Chiara and Luongo, Orlando",
    title = "{Cosmic distance duality after DESI 2024 data release and dark energy evolution}",
    eprint = "2501.15233",
    archivePrefix = "arXiv",
    primaryClass = "astro-ph.CO",
    month = "1",
    year = "2025"
}

@article{Fernandez-Garcia:2025vnb,
    author = "Fern\'andez-Garc\'\i{}a, E. and others",
    title = "{Missing Components in \ensuremath{\Lambda}CDM from DESI Y1 BAO Measurements: Insights from Redshift Remapping}",
    eprint = "2503.22469",
    archivePrefix = "arXiv",
    primaryClass = "astro-ph.CO",
    reportNumber = "FERMILAB-PUB-25-0246-PPD",
    month = "3",
    year = "2025"
}

@article{Bassett:2003zw,
    author = "Bassett, Bruce A.",
    title = "{Cosmic acceleration vs axion - photon mixing}",
    eprint = "astro-ph/0311495",
    archivePrefix = "arXiv",
    doi = "10.1086/383520",
    journal = "Astrophys. J.",
    volume = "607",
    pages = "661--664",
    year = "2004"
}

@article{Schoneberg:2021qvd,
    author = {Sch\"oneberg, Nils and Franco Abell\'an, Guillermo and P\'erez S\'anchez, Andrea and Witte, Samuel J. and Poulin, Vivian and Lesgourgues, Julien},
    title = "{The H0 Olympics: A fair ranking of proposed models}",
    eprint = "2107.10291",
    archivePrefix = "arXiv",
    primaryClass = "astro-ph.CO",
    doi = "10.1016/j.physrep.2022.07.001",
    journal = "Phys. Rept.",
    volume = "984",
    pages = "1--55",
    year = "2022"
}

@book{Weinberg:2008zzc,
    author = "Weinberg, Steven",
    title = "{Cosmology}",
	publisher = {Oxford university press},
    isbn = "978-0-19-852682-7",
    year = "2008"
}

@article{Froustey:2020mcq,
    author = "Froustey, Julien and Pitrou, Cyril and Volpe, Maria Cristina",
    title = "{Neutrino decoupling including flavour oscillations and primordial nucleosynthesis}",
    eprint = "2008.01074",
    archivePrefix = "arXiv",
    primaryClass = "hep-ph",
    doi = "10.1088/1475-7516/2020/12/015",
    journal = "JCAP",
    volume = "12",
    pages = "015",
    year = "2020"
}

@article{Tegmark:1996bz,
    author = "Tegmark, Max and Taylor, Andy and Heavens, Alan",
    title = "{Karhunen-Loeve eigenvalue problems in cosmology: How should we tackle large data sets?}",
    eprint = "astro-ph/9603021",
    archivePrefix = "arXiv",
    doi = "10.1086/303939",
    journal = "Astrophys. J.",
    volume = "480",
    pages = "22",
    year = "1997"
}

@article{Nesseris:2010ep,
    author = "Nesseris, Savvas and Shafieloo, Arman",
    title = "{A model independent null test on the cosmological constant}",
    eprint = "1004.0960",
    archivePrefix = "arXiv",
    primaryClass = "astro-ph.CO",
    doi = "10.1111/j.1365-2966.2010.17254.x",
    journal = "Mon. Not. Roy. Astron. Soc.",
    volume = "408",
    pages = "1879--1885",
    year = "2010"
}

@article{Tang:2024zkc,
    author = "Tang, Li and Lin, Hai-Nan and Wu, Ying",
    title = "{Cosmic distance duality relation in light of time-delayed strong gravitational lensing*}",
    eprint = "2410.08595",
    archivePrefix = "arXiv",
    primaryClass = "astro-ph.CO",
    doi = "10.1088/1674-1137/ad83a8",
    journal = "Chin. Phys. C",
    volume = "49",
    number = "1",
    pages = "015104",
    year = "2025"
}

@article{Qi:2024acx,
    author = "Qi, Jing-Zhao and Jiang, Yi-Fan and Hou, Wan-Ting and Zhang, Xin",
    title = "{Testing the Cosmic Distance Duality Relation Using Strong Gravitational Lensing Time Delays and Type Ia Supernovae}",
    eprint = "2407.07336",
    archivePrefix = "arXiv",
    primaryClass = "astro-ph.CO",
    doi = "10.3847/1538-4357/ad9de4",
    journal = "Astrophys. J.",
    volume = "979",
    number = "1",
    pages = "2",
    year = "2025"
}

@article{Holanda:2012ia,
    author = "Holanda, R. F. L. and Carvalho, J. C. and Alcaniz, J. S.",
    title = "{Model-independent constraints on the cosmic opacity}",
    eprint = "1207.1694",
    archivePrefix = "arXiv",
    primaryClass = "astro-ph.CO",
    doi = "10.1088/1475-7516/2013/04/027",
    journal = "JCAP",
    volume = "04",
    pages = "027",
    year = "2013"
}

@article{Menard:2009yb,
    author = "Menard, Brice and Scranton, Ryan and Fukugita, Masataka and Richards, Gordon",
    title = "{Measuring the galaxy-mass and galaxy-dust correlations through magnification and reddening}",
    eprint = "0902.4240",
    archivePrefix = "arXiv",
    primaryClass = "astro-ph.CO",
    doi = "10.1111/j.1365-2966.2010.16486.x",
    journal = "Mon. Not. Roy. Astron. Soc.",
    volume = "405",
    pages = "1025--1039",
    year = "2010"
}

@article{Gahlaut:2025lhv,
    author = "Gahlaut, Savita",
    title = "{Model\textemdash{}Independent Probe of Cosmic Distance Duality Relation}",
    eprint = "2501.15086",
    archivePrefix = "arXiv",
    primaryClass = "gr-qc",
    doi = "10.1088/1674-4527/adae45",
    journal = "Res. Astron. Astrophys.",
    volume = "25",
    number = "2",
    pages = "025019",
    year = "2025"
}

@article{Ioffe:2015ovl,
    author = "Ioffe, Sergey and Szegedy, Christian",
    title = "{Batch Normalization: Accelerating Deep Network Training by Reducing  Internal Covariate Shift}",
    eprint = "1502.03167",
    archivePrefix = "arXiv",
    primaryClass = "cs.LG",
    month = "2",
    year = "2015"
}

@article{Brieden:2022heh,
    author = "Brieden, Samuel and Gil-Mar\'\i{}n, H\'ector and Verde, Licia",
    title = "{A tale of two (or more) h's}",
    eprint = "2212.04522",
    archivePrefix = "arXiv",
    primaryClass = "astro-ph.CO",
    doi = "10.1088/1475-7516/2023/04/023",
    journal = "JCAP",
    volume = "04",
    pages = "023",
    year = "2023"
}

@article{Cardone:2012vd,
    author = "Cardone, Vincenzo F. and Spiro, Susanna and Hook, Isobel and Scaramella, Roberto",
    title = "{Testing the distance duality relation with present and future data}",
    eprint = "1205.1908",
    archivePrefix = "arXiv",
    primaryClass = "astro-ph.CO",
    doi = "10.1103/PhysRevD.85.123510",
    journal = "Phys. Rev. D",
    volume = "85",
    pages = "123510",
    year = "2012"
}

@article{Ma:2016bjt,
    author = "Ma, Cong and Corasaniti, Pier-Stefano",
    title = "{Statistical Test of Distance\textendash{}Duality Relation with Type Ia Supernovae and Baryon Acoustic Oscillations}",
    eprint = "1604.04631",
    archivePrefix = "arXiv",
    primaryClass = "astro-ph.CO",
    doi = "10.3847/1538-4357/aac88f",
    journal = "Astrophys. J.",
    volume = "861",
    number = "2",
    pages = "124",
    year = "2018"
}

@article{Xu:2022zlm,
    author = "Xu, Bing and Wang, Zhenzhen and Zhang, Kaituo and Huang, Qihong and Zhang, Jianjian",
    title = "{Model-independent Test for the Cosmic Distance\textendash{}Duality Relation with Pantheon and eBOSS DR16 Quasar Sample}",
    eprint = "2212.00269",
    archivePrefix = "arXiv",
    primaryClass = "astro-ph.CO",
    doi = "10.3847/1538-4357/ac9793",
    journal = "Astrophys. J.",
    volume = "939",
    number = "2",
    pages = "115",
    year = "2022"
}

@article{Yang:2024icv,
    author = "Yang, Fan and Fu, Xiangyun and Xu, Bing and Zhang, Kaituo and Huang, Yang and Yang, Ying",
    title = "{Testing the cosmic distance duality relation using Type Ia supernovae and radio quasars through model-independent methods}",
    eprint = "2407.05559",
    archivePrefix = "arXiv",
    primaryClass = "astro-ph.CO",
    month = "7",
    year = "2024"
}

@article{Holanda:2011hh,
    author = "Holanda, R. F. L. and Lima, J. A. S. and Ribeiro, M. B.",
    title = "{Probing the Cosmic Distance Duality Relation with the Sunyaev-Zeldovich Effect, X-rays Observations and Supernovae Ia}",
    eprint = "1104.3753",
    archivePrefix = "arXiv",
    primaryClass = "astro-ph.CO",
    doi = "10.1051/0004-6361/201118343",
    journal = "Astron. Astrophys.",
    volume = "538",
    pages = "A131",
    year = "2012"
}

@article{Goncalves:2011ha,
    author = "Goncalves, R. S. and Holanda, R. F. L. and Alcaniz, J. S.",
    title = "{Testing the cosmic distance duality with X-ray gas mass fraction and supernovae data}",
    eprint = "1109.2790",
    archivePrefix = "arXiv",
    primaryClass = "astro-ph.CO",
    doi = "10.1111/j.1745-3933.2011.01192.x",
    journal = "Mon. Not. Roy. Astron. Soc.",
    volume = "420",
    pages = "L43--L47",
    year = "2012"
}

@article{Holanda:2010vb,
    author = "Holanda, R. F. L. and Lima, J. A. S. and Ribeiro, M. B.",
    title = "{Testing the Distance-Duality Relation with Galaxy Clusters and Type Ia Supernovae}",
    eprint = "1005.4458",
    archivePrefix = "arXiv",
    primaryClass = "astro-ph.CO",
    doi = "10.1088/2041-8205/722/2/L233",
    journal = "Astrophys. J. Lett.",
    volume = "722",
    pages = "L233--L237",
    year = "2010"
}

@article{Beutler:2011hx,
    author = "Beutler, Florian and Blake, Chris and Colless, Matthew and Jones, D. Heath and Staveley-Smith, Lister and Campbell, Lachlan and Parker, Quentin and Saunders, Will and Watson, Fred",
    title = "{The 6dF Galaxy Survey: Baryon Acoustic Oscillations and the Local Hubble Constant}",
    eprint = "1106.3366",
    archivePrefix = "arXiv",
    primaryClass = "astro-ph.CO",
    doi = "10.1111/j.1365-2966.2011.19250.x",
    journal = "Mon. Not. Roy. Astron. Soc.",
    volume = "416",
    pages = "3017--3032",
    year = "2011"
}

@article{Xu:2012hg,
    author = "Xu, Xiaoying and Padmanabhan, Nikhil and Eisenstein, Daniel J. and Mehta, Kushal T. and Cuesta, Antonio J.",
    title = "{A 2\% Distance to z=0.35 by Reconstructing Baryon Acoustic Oscillations - II: Fitting Techniques}",
    eprint = "1202.0091",
    archivePrefix = "arXiv",
    primaryClass = "astro-ph.CO",
    doi = "10.1111/j.1365-2966.2012.21573.x",
    journal = "Mon. Not. Roy. Astron. Soc.",
    volume = "427",
    pages = "2146",
    year = "2012"
}

@article{Blake:2012pj,
    author = "Blake, Chris and others",
    title = "{The WiggleZ Dark Energy Survey: Joint measurements of the expansion and growth history at z \ensuremath{<} 1}",
    eprint = "1204.3674",
    archivePrefix = "arXiv",
    primaryClass = "astro-ph.CO",
    doi = "10.1111/j.1365-2966.2012.21473.x",
    journal = "Mon. Not. Roy. Astron. Soc.",
    volume = "425",
    pages = "405--414",
    year = "2012"
}

@article{Ross:2014qpa,
    author = "Ross, Ashley J. and Samushia, Lado and Howlett, Cullan and Percival, Will J. and Burden, Angela and Manera, Marc",
    title = "{The clustering of the SDSS DR7 main Galaxy sample \textendash{} I. A 4 per cent distance measure at $z = 0.15$}",
    eprint = "1409.3242",
    archivePrefix = "arXiv",
    primaryClass = "astro-ph.CO",
    doi = "10.1093/mnras/stv154",
    journal = "Mon. Not. Roy. Astron. Soc.",
    volume = "449",
    number = "1",
    pages = "835--847",
    year = "2015"
}

@article{BOSS:2015npt,
    author = "Gil-Mar\'\i{}n, H\'ector and others",
    collaboration = "BOSS",
    title = "{The clustering of galaxies in the SDSS-III Baryon Oscillation Spectroscopic Survey: RSD measurement from the LOS-dependent power spectrum of DR12 BOSS galaxies}",
    eprint = "1509.06386",
    archivePrefix = "arXiv",
    primaryClass = "astro-ph.CO",
    doi = "10.1093/mnras/stw1096",
    journal = "Mon. Not. Roy. Astron. Soc.",
    volume = "460",
    number = "4",
    pages = "4188--4209",
    year = "2016"
}

@article{DES:2017rfo,
    author = "Abbott, T. M. C. and others",
    collaboration = "DES",
    title = "{Dark Energy Survey Year 1 Results: Measurement of the Baryon Acoustic Oscillation scale in the distribution of galaxies to redshift 1}",
    eprint = "1712.06209",
    archivePrefix = "arXiv",
    primaryClass = "astro-ph.CO",
    reportNumber = "FERMILAB-PUB-17-586-A-AD-AE-SCD",
    doi = "10.1093/mnras/sty3351",
    journal = "Mon. Not. Roy. Astron. Soc.",
    volume = "483",
    number = "4",
    pages = "4866--4883",
    year = "2019"
}

@article{eBOSS:2017tey,
    author = "Bautista, Julian E. and others",
    collaboration = "eBOSS",
    title = "{The SDSS-IV extended Baryon Oscillation Spectroscopic Survey: Baryon Acoustic Oscillations at redshift of 0.72 with the DR14 Luminous Red Galaxy Sample}",
    eprint = "1712.08064",
    archivePrefix = "arXiv",
    primaryClass = "astro-ph.CO",
    doi = "10.3847/1538-4357/aacea5",
    journal = "Astrophys. J.",
    volume = "863",
    pages = "110",
    year = "2018"
}

@article{eBOSS:2017cqx,
    author = "Ata, Metin and others",
    collaboration = "eBOSS",
    title = "{The clustering of the SDSS-IV extended Baryon Oscillation Spectroscopic Survey DR14 quasar sample: first measurement of baryon acoustic oscillations between redshift 0.8 and 2.2}",
    eprint = "1705.06373",
    archivePrefix = "arXiv",
    primaryClass = "astro-ph.CO",
    doi = "10.1093/mnras/stx2630",
    journal = "Mon. Not. Roy. Astron. Soc.",
    volume = "473",
    number = "4",
    pages = "4773--4794",
    year = "2018"
}

@article{etherington_lx_1933,
	title = {{LX}. \textit{{On} the definition of distance in general relativity}},
	volume = {15},
	issn = {1941-5982, 1941-5990},
	url = {http://www.tandfonline.com/doi/abs/10.1080/14786443309462220},
	doi = {10.1080/14786443309462220},
	language = {en},
	number = {100},
	urldate = {2025-02-14},
	journal = {The London, Edinburgh, and Dublin Philosophical Magazine and Journal of Science},
	author = {Etherington, I.M.H.},
	month = apr,
	year = {1933},
	pages = {761--773},
}

@article{Bogdanos:2009ib,
    author = "Bogdanos, C. and Nesseris, Savvas",
    title = "{Genetic Algorithms and Supernovae Type Ia Analysis}",
    eprint = "0903.2805",
    archivePrefix = "arXiv",
    primaryClass = "astro-ph.CO",
    doi = "10.1088/1475-7516/2009/05/006",
    journal = "JCAP",
    volume = "05",
    pages = "006",
    year = "2009"
}

@article{Euclid:2021cfn,
    author = "Martinelli, M. and others",
    collaboration = "Euclid",
    title = "{Euclid: Constraining dark energy coupled to electromagnetism using astrophysical and laboratory data}",
    eprint = "2105.09746",
    archivePrefix = "arXiv",
    primaryClass = "astro-ph.CO",
    reportNumber = "IFT-UAM/CSIC-21-60",
    doi = "10.1051/0004-6361/202141353",
    journal = "Astron. Astrophys.",
    volume = "654",
    pages = "A148",
    year = "2021"
}

@article{Favale:2024sdq,
    author = "Favale, Arianna and G\'omez-Valent, Adri\`a and Migliaccio, Marina",
    title = "{Quantification of 2D vs 3D BAO tension using SNIa as a redshift interpolator and test of the Etherington relation}",
    eprint = "2405.12142",
    archivePrefix = "arXiv",
    primaryClass = "astro-ph.CO",
    doi = "10.1016/j.physletb.2024.139027",
    journal = "Phys. Lett. B",
    volume = "858",
    pages = "139027",
    year = "2024"
}

@article{Nesseris:2013bia,
    author = "Nesseris, Savvas and Garc\'\i{}a-Bellido, Juan",
    title = "{Comparative analysis of model-independent methods for exploring the nature of dark energy}",
    eprint = "1306.4885",
    archivePrefix = "arXiv",
    primaryClass = "astro-ph.CO",
    reportNumber = "IFT-UAM-CSIC-13-073",
    doi = "10.1103/PhysRevD.88.063521",
    journal = "Phys. Rev. D",
    volume = "88",
    number = "6",
    pages = "063521",
    year = "2013"
}

@article{Sapone:2014nna,
    author = "Sapone, Domenico and Majerotto, Elisabetta and Nesseris, Savvas",
    title = "{Curvature versus distances: Testing the FLRW cosmology}",
    eprint = "1402.2236",
    archivePrefix = "arXiv",
    primaryClass = "astro-ph.CO",
    reportNumber = "IFT-UAM-CSIC-14-094",
    doi = "10.1103/PhysRevD.90.023012",
    journal = "Phys. Rev. D",
    volume = "90",
    number = "2",
    pages = "023012",
    year = "2014"
}

@article{Arjona:2019fwb,
    author = "Arjona, Rub\'en and Nesseris, Savvas",
    title = "{What can Machine Learning tell us about the background expansion of the Universe?}",
    eprint = "1910.01529",
    archivePrefix = "arXiv",
    primaryClass = "astro-ph.CO",
    reportNumber = "IFT-UAM/CSIC-19-130",
    doi = "10.1103/PhysRevD.101.123525",
    journal = "Phys. Rev. D",
    volume = "101",
    number = "12",
    pages = "123525",
    year = "2020"
}

@article{eBOSS:2019qwo,
    author = "Blomqvist, Michael and others",
    collaboration = "eBOSS",
    title = "{Baryon acoustic oscillations from the cross-correlation of Ly$\alpha$ absorption and quasars in eBOSS DR14}",
    eprint = "1904.03430",
    archivePrefix = "arXiv",
    primaryClass = "astro-ph.CO",
    doi = "10.1051/0004-6361/201935641",
    journal = "Astron. Astrophys.",
    volume = "629",
    pages = "A86",
    year = "2019"
}

@article{Tutusaus:2023cms,
    author = "Tutusaus, Isaac and Kunz, Martin and Favre, L{\'e}o",
    title = "{Solving the Hubble tension at intermediate redshifts with dynamical dark energy}",
    eprint = "2311.16862",
    archivePrefix = "arXiv",
    primaryClass = "astro-ph.CO",
    month = "11",
    year = "2023"
}

@article{SDSS:2000hjo,
    author = "York, Donald G. and others",
    collaboration = "SDSS",
    title = "{The Sloan Digital Sky Survey: Technical Summary}",
    eprint = "astro-ph/0006396",
    archivePrefix = "arXiv",
    reportNumber = "FERMILAB-PUB-01-319-A",
    doi = "10.1086/301513",
    journal = "Astron. J.",
    volume = "120",
    pages = "1579--1587",
    year = "2000"
}

@article{Wang:2024rxm,
    author = "Wang, Min and Fu, Xiangyun and Xu, Bing and Huang, Yang and Yang, Ying and Lu, Zhenyan",
    title = "{Testing the cosmic distance duality relation with Type Ia supernova and transverse BAO measurements}",
    eprint = "2407.12250",
    archivePrefix = "arXiv",
    primaryClass = "astro-ph.CO",
    doi = "10.1140/epjc/s10052-024-13049-1",
    journal = "Eur. Phys. J. C",
    volume = "84",
    number = "7",
    pages = "702",
    year = "2024"
}

@article{DESI:2024mwx,
    author = "Adame, A. G. and others",
    collaboration = "DESI",
    title = "{DESI 2024 VI: cosmological constraints from the measurements of baryon acoustic oscillations}",
    eprint = "2404.03002",
    archivePrefix = "arXiv",
    primaryClass = "astro-ph.CO",
    reportNumber = "FERMILAB-PUB-24-0154-PPD",
    doi = "10.1088/1475-7516/2025/02/021",
    journal = "JCAP",
    volume = "02",
    pages = "021",
    year = "2025"
}

@article{DES:2024jxu,
    author = "Abbott, T. M. C. and others",
    collaboration = "DES",
    title = "{The Dark Energy Survey: Cosmology Results with \ensuremath{\sim}1500 New High-redshift Type Ia Supernovae Using the Full 5 yr Data Set}",
    eprint = "2401.02929",
    archivePrefix = "arXiv",
    primaryClass = "astro-ph.CO",
    reportNumber = "FERMILAB-PUB-23-0821-PPD, DES-2023-805",
    doi = "10.3847/2041-8213/ad6f9f",
    journal = "Astrophys. J. Lett.",
    volume = "973",
    number = "1",
    pages = "L14",
    year = "2024"
}

@article{Planck:2018vyg,
    author = "Aghanim, N. and others",
    collaboration = "Planck",
    title = "{Planck 2018 results. VI. Cosmological parameters}",
    eprint = "1807.06209",
    archivePrefix = "arXiv",
    primaryClass = "astro-ph.CO",
    doi = "10.1051/0004-6361/201833910",
    journal = "Astron. Astrophys.",
    volume = "641",
    pages = "A6",
    year = "2020",
    note = "[Erratum: Astron.Astrophys. 652, C4 (2021)]"
}

@ARTICLE{2021PhRvD.103h3533A,
       author = {{Alam}, Shadab and {Aubert}, Marie and {Avila}, Santiago and {Balland}, Christophe and {Bautista}, Julian E. and {Bershady}, Matthew A. and {Bizyaev}, Dmitry and {Blanton}, Michael R. and {Bolton}, Adam S. and {Bovy}, Jo and {Brinkmann}, Jonathan and {Brownstein}, Joel R. and {Burtin}, Etienne and {Chabanier}, Sol{\`e}ne and {Chapman}, Michael J. and {Choi}, Peter Doohyun and {Chuang}, Chia-Hsun and {Comparat}, Johan and {Cousinou}, Marie-Claude and {Cuceu}, Andrei and {Dawson}, Kyle S. and {de la Torre}, Sylvain and {de Mattia}, Arnaud and {Agathe}, Victoria de Sainte and {des Bourboux}, H{\'e}lion du Mas and {Escoffier}, Stephanie and {Etourneau}, Thomas and {Farr}, James and {Font-Ribera}, Andreu and {Frinchaboy}, Peter M. and {Fromenteau}, Sebastien and {Gil-Mar{\'\i}n}, H{\'e}ctor and {Le Goff}, Jean-Marc and {Gonzalez-Morales}, Alma X. and {Gonzalez-Perez}, Violeta and {Grabowski}, Kathleen and {Guy}, Julien and {Hawken}, Adam J. and {Hou}, Jiamin and {Kong}, Hui and {Parker}, James and {Klaene}, Mark and {Kneib}, Jean-Paul and {Lin}, Sicheng and {Long}, Daniel and {Lyke}, Brad W. and {de la Macorra}, Axel and {Martini}, Paul and {Masters}, Karen and {Mohammad}, Faizan G. and {Moon}, Jeongin and {Mueller}, Eva-Maria and {Mu{\~n}oz-Guti{\'e}rrez}, Andrea and {Myers}, Adam D. and {Nadathur}, Seshadri and {Neveux}, Richard and {Newman}, Jeffrey A. and {Noterdaeme}, Pasquier and {Oravetz}, Audrey and {Oravetz}, Daniel and {Palanque-Delabrouille}, Nathalie and {Pan}, Kaike and {Paviot}, Romain and {Percival}, Will J. and {P{\'e}rez-R{\`a}fols}, Ignasi and {Petitjean}, Patrick and {Pieri}, Matthew M. and {Prakash}, Abhishek and {Raichoor}, Anand and {Ravoux}, Corentin and {Rezaie}, Mehdi and {Rich}, James and {Ross}, Ashley J. and {Rossi}, Graziano and {Ruggeri}, Rossana and {Ruhlmann-Kleider}, Vanina and {S{\'a}nchez}, Ariel G. and {S{\'a}nchez}, F. Javier and {S{\'a}nchez-Gallego}, Jos{\'e} R. and {Sayres}, Conor and {Schneider}, Donald P. and {Seo}, Hee-Jong and {Shafieloo}, Arman and {Slosar}, An{\v{z}}e and {Smith}, Alex and {Stermer}, Julianna and {Tamone}, Amelie and {Tinker}, Jeremy L. and {Tojeiro}, Rita and {Vargas-Maga{\~n}a}, Mariana and {Variu}, Andrei and {Wang}, Yuting and {Weaver}, Benjamin A. and {Weijmans}, Anne-Marie and {Y{\`e}che}, Christophe and {Zarrouk}, Pauline and {Zhao}, Cheng and {Zhao}, Gong-Bo and {Zheng}, Zheng},
        title = "{Completed SDSS-IV extended Baryon Oscillation Spectroscopic Survey: Cosmological implications from two decades of spectroscopic surveys at the Apache Point Observatory}",
      journal = "Phys. Rev. D",
         year = 2021,
        month = apr,
       volume = {103},
       number = {8},
          eid = {083533},
        pages = {083533},
          doi = {10.1103/PhysRevD.103.083533},
archivePrefix = {arXiv},
       eprint = {2007.08991},
 primaryClass = {astro-ph.CO},
       adsurl = {https://ui.adsabs.harvard.edu/abs/2021PhRvD.103h3533A}
}

@article{Shafer:2015kda,
    author = "Shafer, Daniel L.",
    title = "{Robust model comparison disfavors power law cosmology}",
    eprint = "1502.05416",
    archivePrefix = "arXiv",
    primaryClass = "astro-ph.CO",
    doi = "10.1103/PhysRevD.91.103516",
    journal = "Phys. Rev. D",
    volume = "91",
    number = "10",
    pages = "103516",
    year = "2015"
}

@article{Blanchard:2022xkk,
    author = "Blanchard, Alain and H{\'e}loret, Jean-Yves and Ili{\'c}, St{\'e}phane and Lamine, Brahim and Tutusaus, Isaac",
    title = "{$\Lambda$CDM is alive and well}",
    eprint = "2205.05017",
    archivePrefix = "arXiv",
    primaryClass = "astro-ph.CO",
    doi = "10.33232/001c.117170",
    journal = "Open J. Astrophys.",
    volume = "7",
    pages = "117170",
    year = "2024"
}

@article{10.1093/mnras/stu523,
    author = {Anderson, Lauren and Aubourg, {\'E}ric and Bailey, Stephen and Beutler, Florian and Bhardwaj, Vaishali and Blanton, Michael and Bolton, Adam S. and Brinkmann, J. and Brownstein, Joel R. and Burden, Angela and Chuang, Chia-Hsun and Cuesta, Antonio J. and Dawson, Kyle S. and Eisenstein, Daniel J. and Escoffier, Stephanie and Gunn, James E. and Guo, Hong and Ho, Shirley and Honscheid, Klaus and Howlett, Cullan and Kirkby, David and Lupton, Robert H. and Manera, Marc and Maraston, Claudia and McBride, Cameron K. and Mena, Olga and Montesano, Francesco and Nichol, Robert C. and Nuza, Sebastián E. and Olmstead, Matthew D. and Padmanabhan, Nikhil and Palanque-Delabrouille, Nathalie and Parejko, John and Percival, Will J. and Petitjean, Patrick and Prada, Francisco and Price-Whelan, Adrian M. and Reid, Beth and Roe, Natalie A. and Ross, Ashley J. and Ross, Nicholas P. and Sabiu, Cristiano G. and Saito, Shun and Samushia, Lado and Sánchez, Ariel G. and Schlegel, David J. and Schneider, Donald P. and Scoccola, Claudia G. and Seo, Hee-Jong and Skibba, Ramin A. and Strauss, Michael A. and Swanson, Molly E. C. and Thomas, Daniel and Tinker, Jeremy L. and Tojeiro, Rita and Magaña, Mariana Vargas and Verde, Licia and Wake, David A. and Weaver, Benjamin A. and Weinberg, David H. and White, Martin and Xu, Xiaoying and Yèche, Christophe and Zehavi, Idit and Zhao, Gong-Bo},
    title = "{The clustering of galaxies in the SDSS-III Baryon Oscillation Spectroscopic Survey: baryon acoustic oscillations in the Data Releases 10 and 11 Galaxy samples}",
    journal = {Monthly Notices of the Royal Astronomical Society},
    volume = {441},
    number = {1},
    pages = {24-62},
    year = {2014},
    month = {04},
    issn = {0035-8711},
    doi = {10.1093/mnras/stu523},
    eprint = {https://academic.oup.com/mnras/article-pdf/441/1/24/3007885/stu523.pdf},
}
\bibliographystyle{JHEP}



\end{document}